\documentclass[11pt]{article}
\usepackage{jheppub}

\usepackage[space]{grffile}
\usepackage{pdfpages}

\usepackage{slashed}
\usepackage{graphicx}
\usepackage{amsmath,amssymb,graphicx} 
\usepackage{gensymb} % degree
\usepackage{epsf,color}
\usepackage[dvipsnames]{xcolor}
\usepackage{hyperref}
\usepackage{cancel}
\usepackage{framed}
\usepackage{hyperref}
\usepackage{float}
\usepackage{multirow}
\usepackage{physics}

\definecolor{nicered}{rgb}{0.5,0.,0.}
\definecolor{nicegreen}{rgb}{0.,0.5,0.}
\definecolor{niceblue}{rgb}{0.,0.,0.5}
\hypersetup{colorlinks,citecolor=nicegreen,linkcolor=nicered,urlcolor=niceblue}
\numberwithin{equation}{section}
\newcommand{\beq}{\begin{equation}}
\newcommand{\eeq}{\end{equation}}
\newcommand{\bea}{\begin{eqnarray}}
\newcommand{\eea}{\end{eqnarray}}
\newcommand{\bear}{\begin{eqnarray}}
\newcommand{\eear}{\end{eqnarray}}

\newcommand{\GeV}{\text{ GeV}}

\newcommand{\ba}{\begin{array}}
\newcommand{\ea}{\end{array}}

\renewcommand{\pb}{{\ensuremath\rm pb}} % \xz{ use \renewcommand here to avoid compile error for \pb }

\newcommand{\etal}{\textsl{et al}\;}

\begin{document}
\preprint{PITT-PACC-2120}

\title{Higgs boson decay to charmonia via $c$-quark fragmentation}

\newcommand{\lae}{\begin{array}{c}\,\sim\vspace{-21pt}\\<
\end{array}}
\newcommand{\gae}{\begin{array}{c}\,\sim\vspace{-21pt}\\>
\end{array}}

\author[a]{Tao Han,}
\affiliation[a]{Pittsburgh Particle Physics, Astrophysics and Cosmology Center,\\ 
Department of Physics and Astronomy, University of Pittsburgh, Pittsburgh, PA 15260, USA}
\author[a]{Adam K. Leibovich,}
\author[a]{Yang Ma,}
\author[a,b]{and Xiao-Ze Tan}
\affiliation[b]{
 School of Physics, Harbin Institute of Technology, Harbin, 150001, People's Republic of China
}

\emailAdd{than@pitt.edu}
\emailAdd{akl2@pitt.edu}
\emailAdd{mayangluon@pitt.edu}
\emailAdd{xz.tan@pitt.edu}
\date{\today}

\abstract{
We calculate the decay branching fractions of the Higgs boson to $J/\psi$ and $\eta_c$ via the charm-quark fragmentation mechanism for the color-singlet and color-octet states in the framework of non-relativistic QCD. 
The decay rates are governed by the charm-quark Yukawa coupling, unlike the decay $H\to J/\psi + \gamma$, which is dominated by the $\gamma^*$-$J/\psi$ mixing. 
We find that the decay branching fractions can be about $2 \times 10^{-5}$ for $H\to c{\bar c}+J/\psi$, and $6 \times 10^{-5}$ for $H\to c{\bar c}+\eta_c$. We comment on the perspective of searching for the Higgs boson to $J/\psi$ transition at the High-Luminosity LHC for testing the charm-quark Yukawa coupling. 
}

\maketitle

\section{Introduction}
\label{sec:intro}

The milestone discovery of Higgs boson ($H$) at the CERN Large Hadron Collider (LHC) in 2012 \cite{Aad:2012tfa,Chatrchyan:2012ufa} was a remarkable success of the Standard Model (SM) of elementary particle physics and the Electroweak Symmetry Breaking mechanism (EWSB). 
The outstanding results of the Higgs boson studies by the ATLAS and CMS collaborations at the LHC are consistent with the SM prediction within the current accuracy for the gauge boson final states of 
$\gamma \gamma, ZZ$ and $WW $ 
\cite{Aaboud:2018ezd,Aad:2015ona,Sirunyan:2018egh}, the third generation of fermions 
for the top quark coupling \cite{Aaboud:2018urx,Sirunyan:2018hoz}, 
and the decays to $\tau \bar{\tau}$ \cite{Aad:2015vsa,Sirunyan:2017khh}
and $b\bar{b}$ \cite{Khachatryan:2016vau,Aaboud:2018zhk,Sirunyan:2018kst,Sirunyan:2018koj,Aad:2019mbh,Noguchi:2019ofj}. 
The Higgs decays to the second generation fermions, however, are much more challenging to observe because of the much weaker Yukawa couplings. While it is promising to observe $H\to \mu^+\mu^-$ with enough integrated luminosity \cite{ATL-PHYS-PUB-2018-054,CMS-PAS-FTR-18-011} because of the clean signature \cite{Plehn:2001qg,Han:2002gp}, the $H\to c\bar c$ channel would be extremely difficult to dig out of the data because of the daunting SM di-jet background at the hadron colliders. 
At present, ATLAS and CMS give the upper limit on Higgs direct decay to charm quark mode of 
$\sigma(p p \rightarrow Z H) \times {\rm BR}(H \rightarrow c \bar{c}) < 2.7\, \pb$  and $ \sigma(VH) \times {\rm BR}(H \to c\bar{c}) < 4.5\, \pb$, which are about $100$ and $70$ times greater than the SM prediction, respectively \cite{Aaboud:2018fhh,CMS:2019hve}. Many dedicated efforts have been made to tackle the problem from different directions \cite{LHCHiggsCrossSectionWorkingGroup:2013rie,Bodwin:2013gca,Bodwin:2014bpa,Perez:2015aoa,Perez:2015lra,Brivio:2015fxa,LHCb-CONF-2016-006,Aaboud:2018fhh,Han:2018juw,Alasfar:2019pmn,Coyle:2019hvs,CMS:2019hve,Aguilar-Saavedra:2020rgo,ATLAS-CONF-2021-021,Carlson:2021tes}, with limited successes. 

A potentially promising method to separate the large QCD background is to consider the decay of the Higgs boson into charmonium associated with a photon, $H\to J/\psi + \gamma$, with effective triggers of $J/\psi \to \mu^+\mu^-$ plus a photon. The branching fraction for this decay mode has been calculated to be ${\rm BR}(H\to J/\psi + \gamma)\simeq 2.8\times 10^{-6}$, within the non-relativistic quantum chromodynamics (NRQCD) framework \cite{Bodwin:2013gca,Bodwin:2014bpa}. Even though the final state from this decay mode is quite distinctive with $J/\psi \to e^+e^-, \mu^+\mu^-$, 
the branching fraction is still rather small, far below the currently accessible limits $3.5 \times 10^{-4}$ and $7.6\times 10^{-4}$, given by ATLAS \cite{Aaboud:2018txb} and CMS \cite{Sirunyan:2018fmm}, respectively.
In addition, the dominant $J/\psi$ production is from the ``vector meson dominance'' contribution 
$\gamma^*\to J/\psi$, rendering this process insensitive to the $Hc\bar c$ Yukawa coupling. 
Other similar processes have been proposed to study the nature Higgs boson \cite{Isidori:2013cla,Kagan:2014ila,Konig:2015qat,Sun:2019cxx}.

To take advantage of the clear decay of $J/\psi$, we study another channel with a charmonium production in the Higgs decay 
\begin{equation} 
	\label{eq:HtoJ}
	H \to c  + \bar{c} + J/\psi\ ({\rm or}\ \eta_c). 
\end{equation}
The dominant contribution to these decay processes is the fragmentation mechanism built upon the initial decay $H \to c\bar{c}$, where the enhancements from the fragmentations result in a relatively high rate.
Within the NRQCD formalism, some diagrams for this process 
have been previously calculated in the literature \cite{Qiao:1998kv,Jiang:2015pah}. 
In this paper, we calculate the full leading-order contributions of the charmonium production in Eq.~(\ref{eq:HtoJ}) via the fragmentation mechanism,  
including both QCD and QED contributions. 
We consider $J/\psi$ and $\eta_c$ production through both the color-singlet and the color-octet Fock states. We find power/logarithmic enhancements to the total decay width due to the fragmentations of the $c$ quark, the photon splitting and the gluon splitting. 
We also properly take into account the running mass effect for the charm quark and the electroweak (EW) correction to the Higgs decay width, which have been often neglected in the literature.
We find that the decay branching fractions can be about $2 \times 10^{-5}$ for $H\to c{\bar c}+J/\psi$, and $6 \times 10^{-5}$ for $H\to c{\bar c}+\eta_c$. 

In the light of the upcoming LHC Run 3 and the High-Luminosity LHC (HL-LHC) \cite{Apollinari:2017cqg}, 
we comment on the perspective on searching for the Higgs boson to $J/\psi$ transition for testing the charm-quark Yukawa coupling, in terms of the signal statistics and the significant background contamination. 
The higher rate and a clean $ J/\psi \to \mu \bar{\mu} $ signal could make this channel searchable by using the existing LHC data or in the future HL-LHC, and potentially improve the sensitivity on testing the Higgs-Charm Yukawa coupling. 

The rest of the paper is organized as follows: In Sec.~II, we give a description on the theoretical formalism and present the calculations for the color-singlet and color-octet states, as well as the EW corrections. In Sec.~III, the phenomenological results and discussions on the perspective of probing the charm-Yukawa coupling are presented. We summarize our findings in Sec.~IV.

%%%%%%%%%%%%%%%%%%%%%%%%%%%%%%%%%%%%%%%%%%
\section{Calculational Formalism}
\label{sec:NRQCD}

NRQCD is an effective theory derived from QCD in the non-relativistic approximation to describe the behavior of bound states made of heavy quark-antiquark pairs ($Q{\bar Q}$) \cite{Bodwin:1994jh}. It is valid when the velocity $v$ of $Q$ (${\bar Q}$) in the $Q{\bar Q}$  center of mass frame is nonrelavistic ($v\ll 1$). In the NRQCD framework, the decay width of the Higgs boson can be factorized as
\begin{eqnarray}
    \Gamma =\sum_\mathbb{N}  {\hat \Gamma}_\mathbb{N}(H \to(Q {\bar Q})[n]+X)\times \langle {\cal O}^h[\mathbb{N}] \rangle,
    \label{eq:Gamma}
\end{eqnarray}
where $\mathbb{N}$ stands for the involved $Q{\bar Q}$ Fock state with quantum numbers $n(^{2S+1}L_J^{\rm [color]})$. 
${\hat \Gamma}_\mathbb{N}$ is the perturbatively calculable short-distance coefficient (SDC), which can be expressed in a differential form 
\begin{equation}
    \label{eq:SDC}
    \dd {\hat \Gamma}_\mathbb{N}=\frac{1}{2 m_H}\frac{|{\cal M}|^2}{\langle {\cal O}^{Q{\bar Q}}\rangle} \dd \Phi_3,
\end{equation}
where $m_H$ is the Higgs boson mass, $\langle {\cal O}^{Q{\bar Q}}\rangle$ is the long-distance matrix element (LDME) for a free $Q{\bar Q}$ pair Fock state. ${\cal M}$ is the perturbative matrix elements from the QCD dynamics and all the spin, color and polarizations are summed over. 
$\dd\Phi_3$ is the $3$-body phase space.
The last factor in Eq.~(\ref{eq:Gamma}), 
${\cal O}^h[\mathbb{N}]$ represents the long-distance matrix elements for an exclusive hadronic quarkonium state $h$, that contains all the non-perturbative hadronization information. 
The leading order color-singlet LDMEs can be related to the wave function at the origin and scale as $v^3$: $\langle {\cal O}^{J/\psi}[^3S_1^{[1]}] \rangle$ and
$\langle {\cal O}^{\eta_c}[^1S_0^{[1]}] \rangle$. Current phenomenological applications for $J/\psi$ and $\eta_c$ also include color-octet LDMEs up to order $v^7$: 
$\langle {\cal O}^{J/\psi}[^3S_1^{[8]}] \rangle$, $\langle {\cal O}^{J/\psi}[^1S_0^{[8]}]\rangle$, $\langle {\cal O}^{J/\psi}[^3P_J^{[8]}] \rangle$, $\langle {\cal O}^{\eta_c}[^3S_1^{[8]}]\rangle$, $\langle {\cal O}^{\eta_c}[^1P_1^{[8]}] \rangle$.
We next present the calculations according their color quantum numbers of singlet and octet. 

%%%%%%%%%%%%%%%%%%%%%%%%%%%%%%%%%%%%%%%%%
\subsection{Color-singlet states} %$^3S_1^{[1]}$, $^1S_0^{[1]}$}
\begin{figure}[tb]
    \centering
    \includegraphics[width=.24\textwidth]{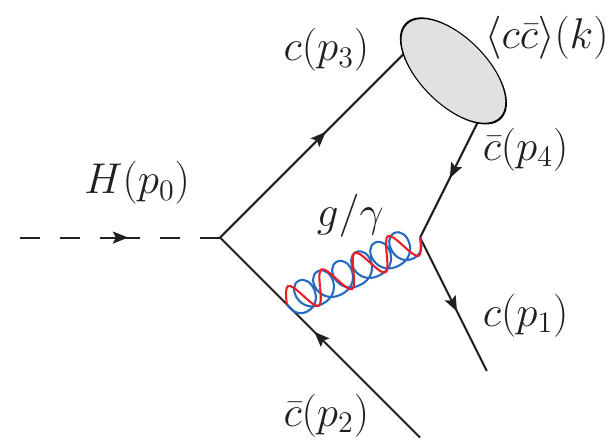}
    \includegraphics[width=.24\textwidth]{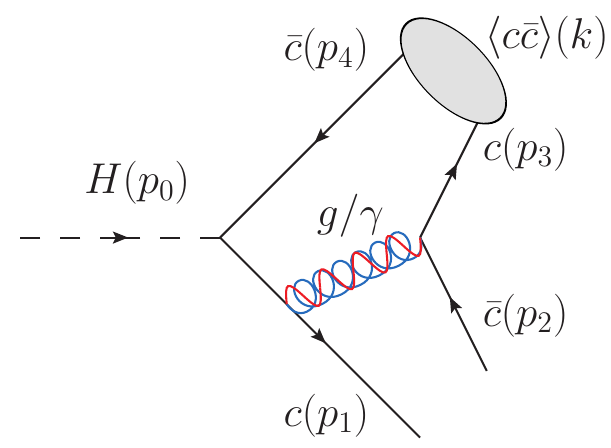}
    \includegraphics[width=.24\textwidth]{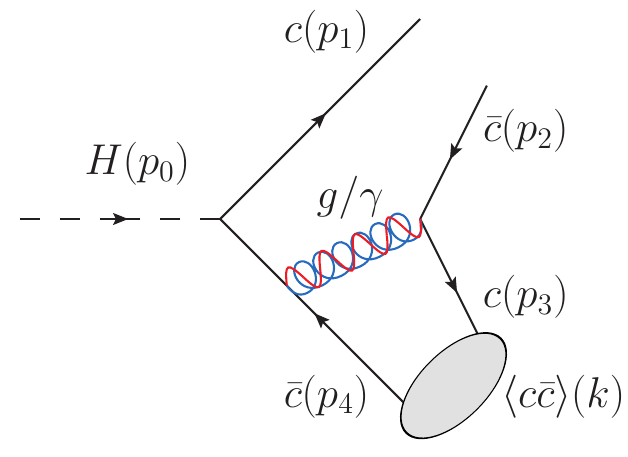}
    \includegraphics[width=.24\textwidth]{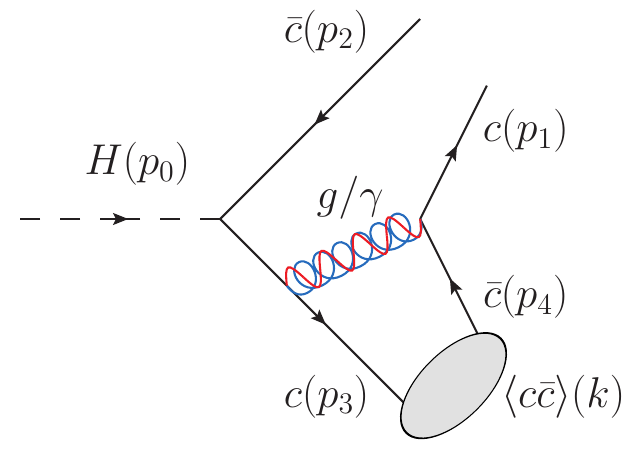}
    \caption{Feynman diagrams for a charmonium Fock state $\langle c {\bar c} \rangle$ production from Higgs decay via charm-quark fragmentation.
    }
    \label{fig:qFrag}
\end{figure}

\begin{figure}[tb]
    \centering
    \includegraphics[width=.25\textwidth]{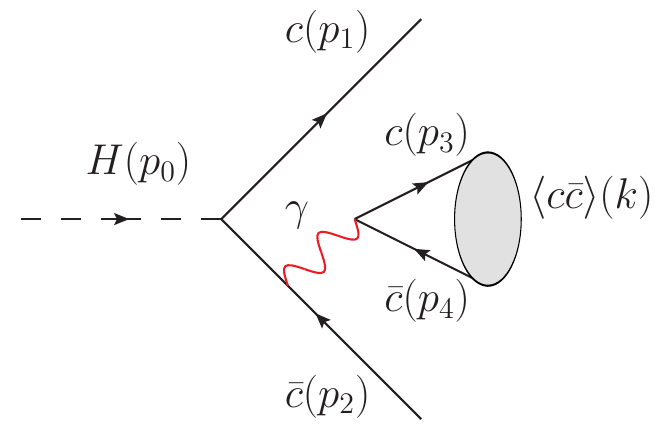}
    \includegraphics[width=.25\textwidth]{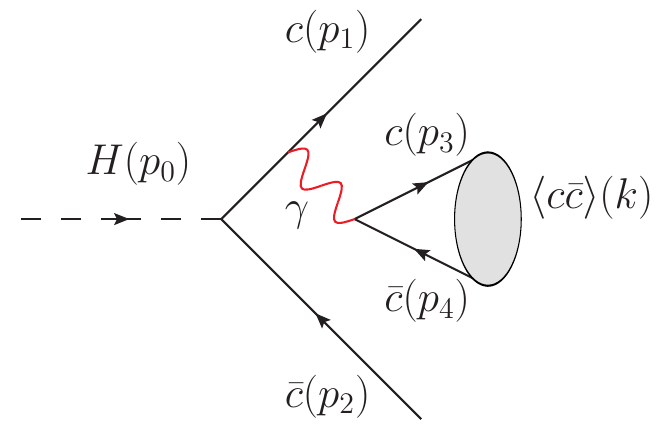}
    \includegraphics[width=.22\textwidth]{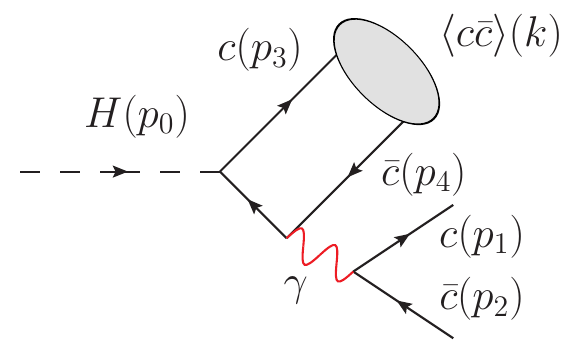}
    \includegraphics[width=.22\textwidth]{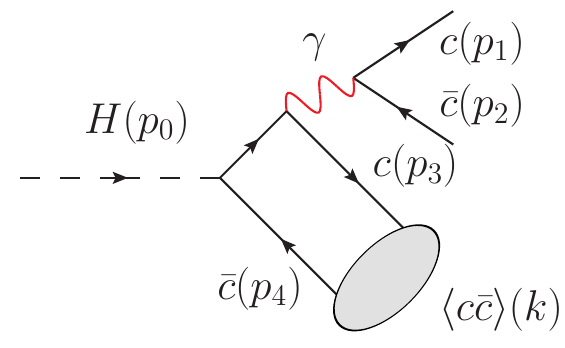}
    %.25 \textwidth for fig a&b and .22 \textwidth for fig c&d make the characters displayed in similar size here.
    \caption{QED Feynman diagrams for color-singlet charmonium state production via 
    $H \to c+{\bar c}+\langle c {\bar c} \rangle$. }
    \label{fig:aFrag}
\end{figure}

\label{sec:CS}
There are two color-singlet Fock states, $^3S_1^{[1]}$ and $^1S_0^{[1]}$, that respectively contributes to $J/\psi$ and $\eta_c$ productions. For the Higgs boson decay to a charmonium bound state $\langle c {\bar c} \rangle$ via the color-singlet Fock states,
\begin{equation}
H(p_0) \to c(p_1)+{\bar c}(p_2)+\langle c {\bar c} \rangle(k),
\end{equation}
the Feynman diagrams are presented in Fig.~\ref{fig:qFrag} for the $g$ and $ \gamma$ contributions, and Fig.~\ref{fig:aFrag} for additional QED only contributions. 

The color-singlet long-distance matrix elements (LDMEs) can be related to the wave function at the origin $R(0)$ by 
\begin{eqnarray}
    \langle {\cal O}^{J/\psi}[^3S_1^{[1]}] \rangle =\frac{3 N_c}{2\pi} |R(0)|^2, ~~~ \langle {\cal O}^{\eta_c}[^1S_0^{[1]}] \rangle= \frac{N_c}{2\pi} |R(0)|^2, \label{eq:CSLDME}
\end{eqnarray}
using the vacuum saturation approximation, valid up to corrections of order $v^4$, where $N_c=3$ is the number of colors. The value of the radial wave function $|R(0)|^2=1.0952 \,{\rm GeV}^3$ can be calculated using the potential models \cite{Eichten:2019hbb}.
% $|R(0)|^2=0.81\,{\rm GeV}^3$ can be calculated using the potential models \cite{Eichten:1995ch}. 
In calculating the SDC as in Eq.~(\ref{eq:SDC}), we have the LDMEs for the free $Q{\bar Q}$ color-singlets
\begin{equation}
    \label{eq:LDME0}
        \langle {\cal O}^{Q{\bar Q}}\rangle =6 N_c ,\  {\text {\rm for }} ^3S_1^{[1]}, \quad
        \langle {\cal O}^{Q{\bar Q}}\rangle =2 N_c ,\  {\text {\rm for }} ^1S_0^{[1]} .
\end{equation}
The Feynman amplitudes have the form
\begin{eqnarray}
    {\cal M} =\frac{y_c}{\sqrt 2} \frac{\delta_{ij}}{\sqrt {N_c}} \sum_{\ell=1}^4 {\bar u}(p_1)\epsilon_\alpha\left.\left[\left(C_F g_s^2+q_c^2 e^2\right){\cal A}_\ell^\alpha + C_A q_c^2 e^2{\cal B}_\ell^\alpha) \right]\right|_{q=0} v(p_2), \label{eq:CSamp}
\end{eqnarray}
where $y_c$ and $q_c$ are the charm-quark Yukawa coupling and the electric charge, respectively. 
$\epsilon_\alpha$ is the polarization vector of the $Q{\bar Q}$ Fock state,\footnote{For a spin-zero state such as $^1S_0^{[1]}$, $\epsilon_\alpha\to 1$ and ${\cal A}^\alpha,\ {\cal B}_\ell^\alpha$ are scalar functions independent of $\alpha$, as given explicitly below.}
$q=p_3-p_4$ is the relative momentum between the constitute quarks $Q$ and ${\bar Q}$, $i$ and $j$ are the color indices of $Q$ and ${\bar Q}$, $C_A=3$, $C_F=4/3$, 
$g_s$ is the strong coupling, and $q_c$ is the charm quark electric charge.  
The dominant contribution ${\cal A}_\ell^\alpha$ is from the ``quark fragmentation mechanism'', which can be read from the Feynman diagrams in Fig.~\ref{fig:qFrag},
\begin{eqnarray}
    &&{\cal A}_1^\alpha=-\frac{\gamma^\nu \Pi_s^\alpha (m_c-\slashed p_1-\slashed p_2-\slashed p_4)\gamma^\nu }{(p_1+p_4)^2((p_1+p_2+p_4)^2-m_c^2)}, \ \ 
    {\cal A}_2^\alpha=-\frac{\gamma^\nu (m_c+\slashed p_1+\slashed p_2+\slashed p_3)\Pi_s^\alpha \gamma^\nu }{(p_2+p_3)^2((p_1+p_2+p_3)^2-m_c^2)},\nonumber\\
    &&{\cal A}_3^\alpha=-\frac{(m_c-\slashed p_2-\slashed p_3-\slashed p_4)\gamma^\nu \Pi_s^\alpha \gamma^\nu }{(p_2+p_3)^2((p_2+p_3+p_4)^2-m_c^2)},\ \ 
    {\cal A}_4^\alpha=-\frac{\gamma^\nu \Pi_s^\alpha \gamma^\nu(m_c+\slashed p_1+\slashed p_3+\slashed p_4) }{(p_1+p_4)^2((p_1+p_3+p_4)^2-m_c^2)},~~\qquad
    \label{eq:CSqFrag}
\end{eqnarray}
where $\Pi_s^\alpha$ is a spin projector for a spin-$s$ state. The pure QED amplitudes ${\cal B}_\ell^\alpha$ can be read off from Fig.~\ref{fig:aFrag},
\begin{eqnarray}
    &&{\cal B}_1^\alpha=\frac{{\rm Tr}[\gamma^\nu \Pi_s^\alpha] (m_c-\slashed p_2-\slashed p_3-\slashed p_4)\gamma^\nu }{(p_3+p_4)^2((p_2+p_3+p_4)^2-m_c^2)}, \ \
   {\cal B}_2^\alpha=\frac{{\rm Tr}[\gamma^\nu \Pi_s^\alpha] \gamma^\nu (m_c+\slashed p_1+\slashed p_3+\slashed p_4) }{(p_3+p_4)^2((p_1+p_3+p_4)^2-m_c^2)},\nonumber \\
    &&{\cal B}_3^\alpha=\frac{\gamma^\nu {\rm Tr}[(m_c-\slashed p_1-\slashed p_2-\slashed p_4) \gamma^\nu \Pi_s^\alpha ]}{(p_1+p_2)^2((p_1+p_2+p_4)^2-m_c^2)},\ \ 
    {\cal B}_4^\alpha=\frac{\gamma^\nu {\rm Tr}[ \gamma^\nu(m_c+\slashed p_1+\slashed p_2+\slashed p_3) \Pi_s^\alpha]}{(p_1+p_2)^2((p_1+p_2+p_3)^2-m_c^2)}.~~\qquad
    \label{eq:CSQED}
\end{eqnarray}
Thanks to the ``single photon fragmentation'' mechanism in Fig.~\ref{fig:aFrag}(a,b), the QED diagrams have a notable enhancement to $^3S_1^{[1]}$ production via their interference with the QCD diagrams. Meanwhile, for CP conservation, the single-photon-fragmentation diagrams are forbidden in $^1S_0^{[1]}$ production.
The spin projectors $\Pi_s^\alpha$ for the outgoing heavy quark pair are given by
\begin{eqnarray}
    &&\Pi_0^\alpha \to \frac{1}{\sqrt {8 m_c^3}}\left( \frac{\slashed k}{2}-\slashed q -m\right)\gamma^5 \left( \frac{\slashed k}{2}+\slashed q + m\right),\nonumber \\
    &&\Pi_1^\alpha=\frac{1}{\sqrt {8 m_c^3}}\left( \frac{\slashed k}{2}-\slashed q -m\right)\gamma^\alpha \left( \frac{\slashed k}{2}+\slashed q + m\right), \label{eq:projector}
\end{eqnarray} 
for spin-$0$ and spin-$1$ states respectively, where $k=p_3+p_4$. 
By substituting Eqs.~(\ref{eq:CSLDME})$-$(\ref{eq:projector}) into Eq.~(\ref{eq:Gamma}), one obtains the decay width of $\Gamma(H \to c{\bar c} + J/\psi (\eta_c))$ through the color-singlet states. The polarization sum formulae are listed in Appendix \ref{Appendix:polsum}.

%%%%%%%%%%%%%%%%%%%%%%%%%%%%%%%%%%%%%%%%%%%
\subsection{Color-octet states}
\label{sec:CO}
A key property of NRQCD is that a quarkonium can also be produced through color-octet Fock states. The color-octet long-distance matrix elements (LDMEs) have to be extracted from fitting the experimental data with the NRQCD calculations. Different fitting strategies result in different values of LDMEs; some of the recent color-octet LDMEs fitting results for the $J/\psi$ production are listed in Table \ref{tab:LDMEs}. 
In our computation, a combined fit of CDF and CMS $J/\psi$ production data for the color-octet LDMEs \cite{Bodwin:2014gia} is employed. 
One reason for choosing this extraction is due to the fact that it relies on high $p_T$ hadronic data. Since the possible factorization issues at small $p_T$ are not present for the Higgs decay, we feel that the extraction in Ref.~\cite{Bodwin:2014gia} is closest to our current interest, and will thus use these as our canonical value for the LDMEs. We also note another merit that the color-octet LDMEs in Ref.~\cite{Bodwin:2014gia} is independent of the value of the wave function at origin.

Based on the heavy quark spin symmetry (HQSS), there exist the following relations
\begin{eqnarray}
  &  \langle {\cal O}^{\eta_c}[^1S_0^{[1,8]}] \rangle &= \frac{1}{3}\langle {\cal O}^{J/\psi}[^3S_1^{[1,8]}] \rangle,~~ 
    \nonumber \\
&    \langle {\cal O}^{\eta_c}[^3S_1^{[8]}] \rangle &= \langle {\cal O}^{J/\psi}[^1S_0^{[8]}] \rangle, \ \ 
    \langle {\cal O}^{\eta_c}[^1P_1^{[8]}] \rangle = 3 \langle {\cal O}^{J/\psi}[^3P_0^{[8]}] \rangle,
\end{eqnarray}
that allow us to relate all the needed LDMEs to
those in Table \ref{tab:LDMEs}.

\begin{table}[tb] 
    \caption{Some fitted numerical values of color-octet long-distance matrix elements (LDMEs) for $J/\psi$ production (in units of $\GeV^3$)}\label{tab:LDMEs}
    \center
    \scalebox{0.9}{
   	 \begin{tabular}{lccc}
   		\hline
   		Reference            & $\langle {\cal O}^{J/\psi}[^1S_0^{[8]}] \rangle$ & $\langle {\cal O}^{J/\psi}[^3S_1^{[8]}] \rangle$ & $\langle {\cal O}^{J/\psi}[^3P_0^{[8]}] \rangle/m_c^2$ \\ \hline
   		G. Bodwin, \etal\cite{Bodwin:2014gia}        & $(9.9\pm2.2)\times10^{-2}$                       & $(1.1\pm1.0)\times10^{-2}$                      & $(4.89\pm4.44)\times 10^{-3}$                           \\
   		K.T. Chao, \etal\cite{Chao:2012iv}      & $(8.9\pm0.98)\times10^{-2}$                      & $(3.0\pm1.2)\times 10^{-3}$                      & $(5.6\pm2.1)\times 10^{-3}$                            \\
   		Y. Feng, \etal\cite{Feng:2018ukp}        & $(5.66\pm4.7)\times 10^{-2}$                     & $(1.77\pm0.58)\times 10^{-3}$                    & $(3.42\pm1.02)\times 10^{-3}$                          \\ \hline
   	\end{tabular}
   }
\end{table}

\begin{figure}[tb]
    \centering
    \includegraphics[width=.25\textwidth]{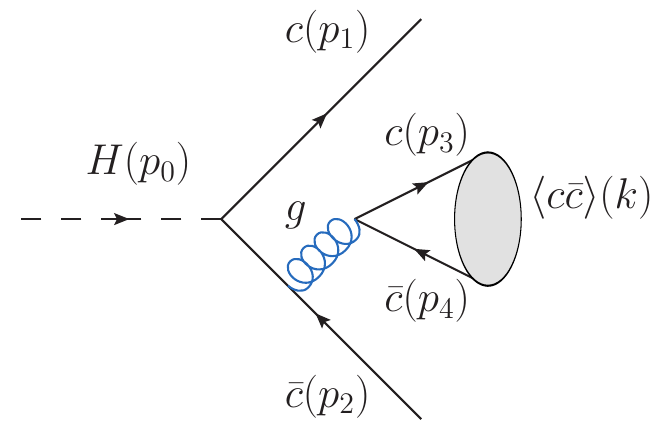}
    \includegraphics[width=.25\textwidth]{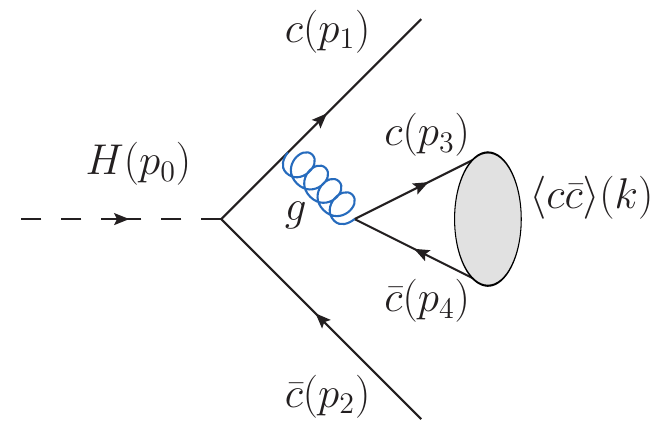}
    \includegraphics[width=.22\textwidth]{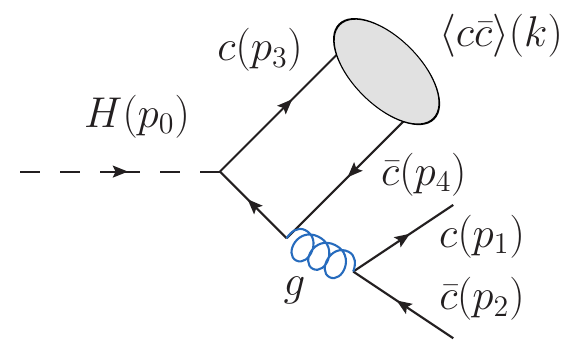}
    \includegraphics[width=.22\textwidth]{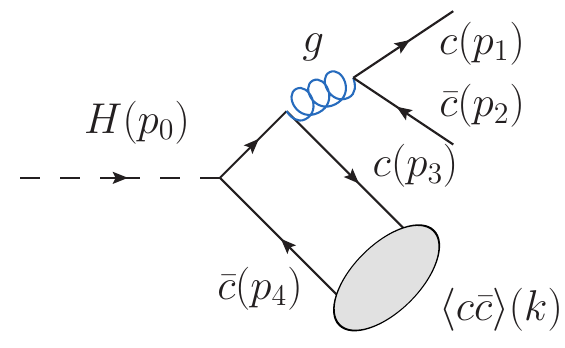}
    %.25 \textwidth for fig a&b and .22 \textwidth for fig c&d make the characters displayed in similar size here.
    \caption{Feynman diagrams for color-octet charmonium state production. (a) and (b) are the single gluon fragmentation to $^3S_1^{[8]}$ state, while (c) and (d) contribute to both $^3S_1^{[8]}$ and $^1P_1^{[8]}$ states. }
    \label{fig:gFrag}
\end{figure}

The short-distance coefficient (SDC) calculation for the color-octet states is similar to those for the color-singlet ones, with the free $Q{\bar Q}$ pair state LDMEs
\begin{eqnarray}
    &&\langle {\cal O}^{Q{\bar Q}}(^1S_0^{[8]})\rangle= (N_c^2-1),~~~\langle {\cal O}^{Q{\bar Q}}(^3S_1^{[8]})\rangle=3 (N_c^2-1) ,\nonumber\\  
    &&\langle {\cal O}^{Q{\bar Q}}(^1P_1^{[8]})\rangle= 3 (N_c^2-1),~\langle {\cal O}^{Q{\bar Q}}(^3P_J^{[8]})\rangle=(2J+1) (N_c^2-1) ,~J=0,\,1,\,2.\label{eq:CO_QQLDME}
\end{eqnarray}
In addition to the Feynman diagrams in Fig.~\ref{fig:qFrag}, there are new QCD Feynman diagrams for the color-octet final states, as shown in Fig.~\ref{fig:gFrag}. The ``single gluon fragmentation'' diagrams in Fig.~\ref{fig:gFrag}(a,b) contribute only to $^3S_1^{[8]}$ and causes it to dominant over the other color-octet states. The Fig.~\ref{fig:gFrag}(c,d) diagrams are non-zero only for the $^3S_1^{[8]}$ and the $^1P_1^{[8]}$ cases due to the CP symmetry. 
Again, following Eq.~(\ref{eq:SDC}), the color-octet Feynman amplitudes can be written as 
\begin{eqnarray}
    {\cal M} =\frac{y_c}{\sqrt 2} {\bar u}(p_1) {\cal M'} v(p_2), \label{eq:COamp}
\end{eqnarray}
where
\begin{eqnarray}
    {\cal M'}=\sqrt{2} \epsilon_\alpha  \left.\left\{ \left[ \left(T^aT^bT^a\right)_{ij} \, g_s^2+T^b_{ij} q_c^2 e^2 \right] \sum_{\ell=1}^4 {\cal A}_\ell^\alpha + \frac{T^b_{ij}}{2}g_s^2 \sum_{\ell=1}^4 {\cal B}_\ell^\alpha) \right\}\right|_{q=0}, \label{eq:COSamp}
\end{eqnarray}
for $s$-wave states (note $\epsilon_\alpha\to 1$ and ${\cal B}_\ell^\alpha\to 0$ for $^1S_0^{[8]}$),
\begin{eqnarray}
    {\cal M'}=\sqrt{2} \epsilon_\beta  \frac{\dd}{\dd q_\beta}\left.\left\{ \left[ \left(T^aT^bT^a\right)_{ij} \, g_s^2+T^b_{ij} q_c^2 e^2 \right] \sum_{\ell=1}^4{\cal A}_\ell^\alpha + \frac{T^b_{ij}}{2}g_s^2 \sum_{\ell=3}^4 {\cal B}_\ell^\alpha) \right\}\right |_{q=0}\label{eq:1p18amp}
\end{eqnarray}
for $^1P_1^{[8]}$, and
\begin{eqnarray}
    {\cal M'}=\sqrt{2} {\cal E}_{\alpha\beta} \frac{\dd}{\dd q_\beta}\left. \left[ \left(T^aT^bT^a\right)_{ij} \, g_s^2+T^b_{ij} q_c^2 e^2 \right]\sum_{\ell=1}^4{\cal A}_\ell^\alpha \right |_{q=0}\label{eq:3pJ8amp}
\end{eqnarray}
for $^3P_J^{[8]}$ ($J=0,\,1,\,2$). The polarization vector and tensor are denoted by $\epsilon_\alpha$ and ${\cal E}_{\alpha\beta}$, and $b$ is for the color of the color-octet $Q{\bar Q}$ Fock state.

A special remark is in order about the color-octet mechanism in the $J/\psi$ and $\eta_c$ production.
As described in Eq.~(\ref{eq:Gamma}), the decay width $\Gamma$ can be factorized into the LDME $\langle {\cal O}^h[\mathbb{N}] \rangle$ and the SDC $\hat{\Gamma}_\mathbb{N}$, following the NRQCD framework. The color-octet LDMEs are in higher orders of $v$ than the color-singlet ones as
\begin{eqnarray}
    &&\frac{\langle {\cal O}^{J/\psi}(^1S_0^{[8]})\rangle}{\langle {\cal O}^{J/\psi}(^3S_1^{[1]})\rangle} \sim {\cal O} (v^3), ~~~\frac{\langle {\cal O}^{J/\psi}(^3S_1^{[8]})\rangle}{\langle {\cal O}^{J/\psi}(^3S_1^{[1]})\rangle} \sim {\cal O} (v^4), ~~~\frac{\langle {\cal O}^{J/\psi}(^3P_J^{[8]})\rangle}{\langle {\cal O}^{J/\psi}(^3S_1^{[1]})\rangle} \sim {\cal O} (v^4),\label{eq:CO_LDME_v} \nonumber \\
    &&\frac{\langle {\cal O}^{\eta_c}(^3S_1^{[8]})\rangle}{\langle {\cal O}^{\eta_c}(^1S_0^{[1]})\rangle} \sim {\cal O} (v^3),
    ~~~\frac{\langle {\cal O}^{\eta_c}(^1P_1^{[8]})\rangle}{\langle {\cal O}^{\eta_c}(^1S_0^{[1]})\rangle} \sim {\cal O} (v^4),
\end{eqnarray}
which naively suppresses the rates to produce $J/\psi$ and $\eta_c$ via the color-octet states. 
The SDCs for different Fock states can be very different since they may include different contributing diagrams and therefore different color structures. We present the color factors of different Feynman diagrams for the color-singlet and color-octet SDCs in Table~\ref{tab:ColorFactor}. 
As shown in the table, the QCD quark fragmentation mechanism (the Feynman diagrams with a gluon propagator in Fig.~\ref{fig:qFrag}) is suppressed in the color-octet productions by a factor of 8.
Among all the color-octet states, $^3S_1^{[8]}$ has the largest SDC, due to both its relatively larger color factor of Fig.~\ref{fig:gFrag} and the large logarithmic single-gluon-fragmentation enhancement from Fig.~\ref{fig:gFrag}(a,b).
For the other color-octet states, {\it i.e.} $^1S_0^{[8]}$, $^1P_1^{[8]}$, and $^3P_J^{[8]}$, the main production process is via charm-quark fragmentation as shown in Fig.~\ref{fig:qFrag}, where the QED diagrams make sizeable contributions via the QCD/QED interference terms because of a large color factor.

\begin{table}[tb]
    \caption{Color factors of different Feynman diagrams for the color-singlet (CS) and color-octet (CO) short-distance coefficients. The pure QCD contribution, pure QED contribution and the QCD/QED interference are represented as QCD, QED, and QCD$\times$QED, respectively.}\label{tab:ColorFactor}
    \center
    \scalebox{0.9}{
    \begin{tabular}{cccccc}
    \hline
       &        & Fig.~\ref{fig:qFrag} &                 & Fig.~\ref{fig:aFrag}   & Fig. ~\ref{fig:gFrag}   \\ \hline
       & QCD    & QED   & QCD$\times$ QED & QED      & QCD      \\ \hline
    CS & $16/9$ & $1$   & $4/3$           & $9$      & -        \\
    CO & $2/9$  & $8$   & $-4/3$          & -        & $2$      \\ \hline
    \end{tabular}
    }
\end{table}

%%%%%%%%%%%%%%%%%%%%%%%%%%%%%%%%%%%%%%%%%
\subsection{Electroweak contributions}
\begin{figure}[tb]
    \centering
    \includegraphics[width=.24\textwidth]{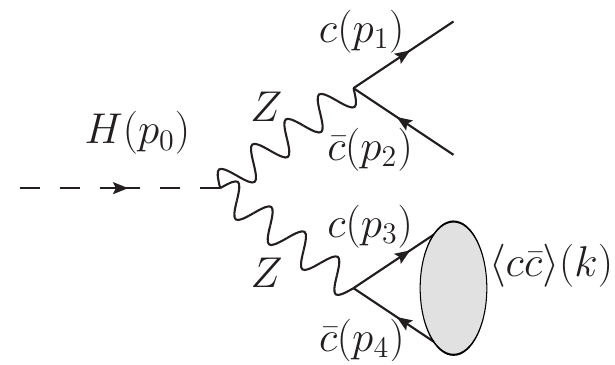}
    \includegraphics[width=.24\textwidth]{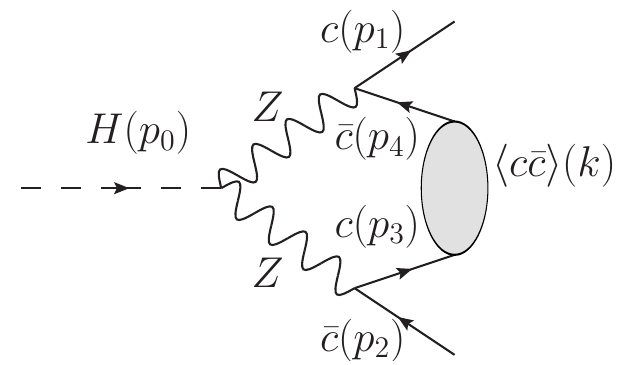}
    \caption{Feynman diagrams for charmonium state production through the $HZZ$ coupling.}
    \label{fig:HZZ}
\end{figure}

Besides the Feynman diagrams in Figs.~\ref{fig:qFrag}, \ref{fig:aFrag} and \ref{fig:gFrag}, we also consider the electroweak (EW) production mechanism through the $HZZ$ coupling, as shown in Fig.~\ref{fig:HZZ}. 
The color factors of these two Feynman diagrams are listed in Table~\ref{tab:ColorFactorHZZ}.
The Feynman diagram in Fig.~\ref{fig:HZZ}(a) could give a sizable correction for the color-singlet states productions for both its relatively larger color factor ($5$ times of the charm quark QCD fragmentation) and the resonance enhancement of the on-shell $Z$ splitting to a pair of free $c{\bar c}$.
For Fig.~\ref{fig:HZZ}(b), one of the two $Z$ propagators could be very closed to $Z$ mass shell with $p_Z^2\leq m_H^2/2-4m_c^2 \simeq (88.34~ {\rm GeV})^2$, so its contribution is also non-negligible. 
Particularly, for the color-octet state production, where only Fig.~\ref{fig:HZZ}(b) exists, the EW correction can be quite large due to nearly on-shell $Z$ enhancement and the relatively larger color factor ($36$ times of the charm quark QCD fragmentation). 
We note this EW contribution possesses contamination to the charm-Yukawa coupling measurement. 
The branching fractions for the $HZZ$ contribution are estimated to be around $6\times 10^{-7}$ for $J/\psi$ production and $3\times 10^{-6}$ for $\eta_c$ production. More detailed numerical comparisons will be shown in the following section.

\begin{figure}[tb]
	\centering
	\includegraphics[width=.24\textwidth]{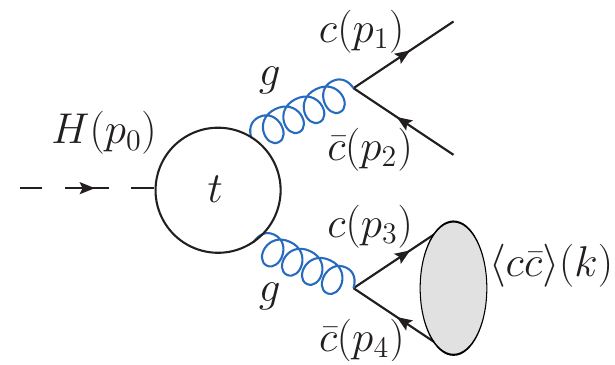}
	\includegraphics[width=.24\textwidth]{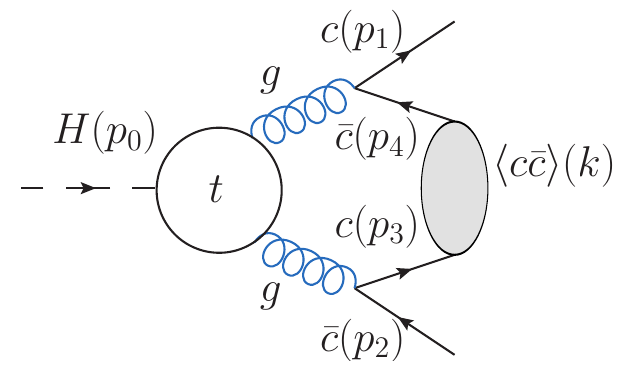}
	\includegraphics[width=.24\textwidth]{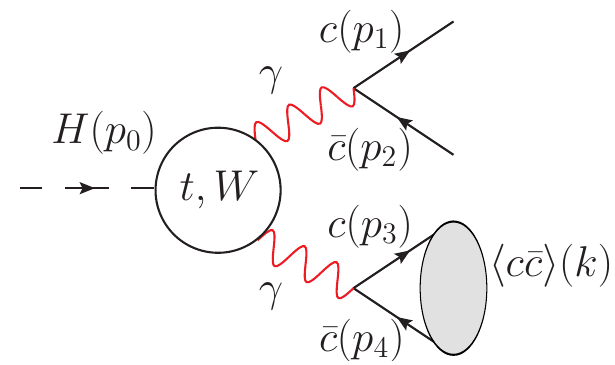}
	\includegraphics[width=.24\textwidth]{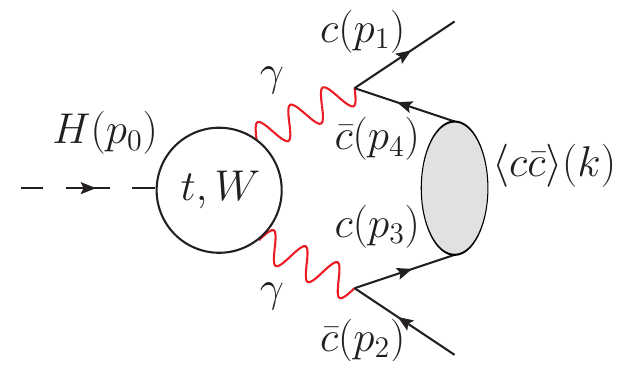}
	\caption{Feynman diagrams for charmonium state production with top-quark and $W$ loop contributions. The gluonic diagram in (a) only contributes to $ ^3 S_1 ^{[8] }$, while the photon one in (c) only contributes to $ ^3 S_1 ^{[1]} $.
	}
	\label{fig:H-loop}
\end{figure}

Before ending this section, one remark is in order. Owing to the large top-quark Yukawa coupling, the Higgs boson decay via the top-quark loop may be substantial. 
The best known example, as the Higgs boson discovery channel, is $gg\to H$ and  
$H\to \gamma\gamma$, 
which would also contribute to the final state of our current interest. We show the contributing Feynman diagrams for $H\to g^* g^*/\gamma^*\gamma^* \to J/\psi + c\bar c$ in Fig.~\ref{fig:H-loop}. 
The branching fraction for the $g^*g^*$ contribution is estimated to be around $2.5\times 10^{-6}$ in the heavy top limit, which is significantly smaller than that from the charm-Yukawa contributions. 
As already noted earlier, the decay $H\to J/\psi + \gamma$ is dominated by the vector meson dominance contribution via $H\to \gamma^*\gamma \to J/\psi + \gamma$.
The photon splitting will contribute to the final state under our consideration $H\to \gamma^*\gamma^* \to J/\psi + c\bar c$. However, it is quite small, less than $2\times 10^{-7}$. We will not discuss those contributions further.

\begin{table}[tb]
    \caption{Color factors of the $HZZ$ diagrams for the color-singlet (CS) and color-octet (CO) short-distance coefficients.}\label{tab:ColorFactorHZZ}
    \center
    \scalebox{0.9}{
    \begin{tabular}{ccc}
    \hline
       & Fig.~\ref{fig:HZZ}(a) & Fig.~\ref{fig:HZZ}(b) \\ \hline
    CS & $9$        & $1$        \\
    CO & -          & $8$        \\ \hline
    \end{tabular}
    }
\end{table}

%%%%%%%%%%%%%%%%%%%%%%%%%%%%%%%%%%%%%
\section{Phenomenological results}

\subsection{Standard Model results}

In our numerical calculations, the SM parameters are taken as
\begin{eqnarray}
    &&1/\alpha=132.5, ~~~\alpha_s(2 m_c)=0.2353,\quad m_c^{\rm pole}=1.500~{\rm GeV},
    ~~~m_c(m_H)=0.6942~{\text {\rm GeV}},\nonumber \\
        &&m_H=125.0~{\text {\rm GeV}},~~~ m_W=80.42~{\text {\rm GeV}},~~~ m_Z=91.19~{\text {\rm GeV}},
        ~~~v =246.2\ {\text {\rm GeV}}, \nonumber
\end{eqnarray}
where the QCD running coupling $\alpha_s(2 m_c)$ and the charm quark running mass $m_c(m_H)$ are obtained by running from $\alpha_s(m_Z)=0.1181$ \cite{Tanabashi:2018oca} and $m_c(3 ~{\rm GeV})=1.012~{\rm GeV}$ \cite{Kuhn:2007vp} at one-loop level.\footnote{The amplitude square can be analytically simplified using \texttt{FeynCalc} \cite{Mertig:1990an,Shtabovenko:2016sxi,Shtabovenko:2020gxv}, the numerical parameters for the SM parameters, $\alpha_s(Q)$ running, and $m_c(Q)$ running are implemented using \texttt{para} \cite{Han:2020uid,para}.} The SM Yukawa coupling at the scale of the Higgs boson mass is 
\begin{eqnarray}
    y_c^{\rm SM}= \frac{{\sqrt 2}m_c(m_H)}{ v} \approx 3.986\times 10^{-3}, \nonumber
\end{eqnarray}
which gives a branching fraction ${\rm BR}(H\to c{\bar c})=2.9\%$, consistent with Ref.~\cite{deFlorian:2016spz}. 

% Yang rewrote this part for discussing the SDC only
\begin{table}[tb]
	\caption{The ratios of the SDCs to their pure QCD values ${\hat\Gamma}_\mathbb{N}/{\hat\Gamma}^{\rm QCD}_\mathbb{N}$. The pure QCD contribution, pure QED contribution, QCD/QED interference, and EW correction are marked as QCD, QED, QCD$\times$QED, and EW, respectively.} 
	\label{tab:SDC}
	\center
	\scalebox{0.9}{
		\begin{tabular}{ccccccccc}
			\hline
			${\hat\Gamma}_\mathbb{N}/{\hat\Gamma}_\mathbb{N}^{\rm QCD}$ & $^1S_0^{[1]}$ & $^3S_1^{[1]}$ & $^1S_0^{[8]}$ & $^3S_1^{[8]}$ & $^1P_1^{[8]}$ & $^3P_0^{[8]}$ & $^3P_1^{[8]}$ & $^3P_2^{[8]}$ \\ \hline
			QCD                                   & 1.0         & 1.0         & 1.0         & 1.0         & 1.0         & 1.0         & 1.0         & 1.0         \\
			QED                                   & $1.1\times10^{-4}$      & $0.077$       & $0.0073$       & $1.1\times10^{-5}$     & $0.0068$       & $0.0073$       & $0.0073$       & $0.0073$       \\
			QCD$\times$QED                        & $0.021$       & $0.14$       & $-0.17$     & $0.0012$      & $-0.15$      & $-0.17$      & $-0.17$      & $-0.17$      \\
			EW                                    & $0.24$       & $0.051$       & $0.28 $       & $2.6\times10^{-4}$      & $1.4$        & $0.29$       & $0.33$       & $1.5$        \\ \hline
		\end{tabular}
	}
\end{table}

The numerical short-distance coefficients (SDCs) can be obtained by substituting the Feynman amplitudes Eq.~(\ref{eq:CSamp}) and Eq.~(\ref{eq:COamp}) into Eq.~(\ref{eq:SDC}). We decompose the SDCs into pure QCD contribution, pure QED contribution, QCD/QED interference, the EW correction, and present the ratios of the SDCs to the corresponding pure QCD values ${\hat\Gamma}_\mathbb{N}/{\hat\Gamma}^{\rm QCD}_\mathbb{N}$ in Table~\ref{tab:SDC}. The QCD diagrams dominate for the SDCs of both the color-singlet states and most of the color-octet states, especially for $^3S_1^{[8]}$.
The QED diagrams introduce sizable corrections mainly via the QCD/QED interference, which affects different Fock states differently:
\begin{itemize}
    \item For $^3S_1^{[1]}$, the QED contribution is enhanced by both the logarithmic enhancement and the large color factor of the single-photon-fragmentation diagrams (Fig.~\ref{fig:aFrag}~(a, b)). Together with the QCD/QED interference, the total QED correction is around $22\%$ compared to the pure QCD contribution.
    \item For $^1S_0^{[1]}$, the Fig.~\ref{fig:aFrag} diagrams are forbidden by CP conservation, leading to the total QED correction of only $2\%$.
    \item For $^3S_1^{[8]}$, the QCD contribution is absolutely dominant for the single-gluon-fragmentation  diagrams (Fig.~\ref{fig:gFrag} (a,b)), for which both the QED and EW corrections are orders of magnitude smaller.
    \item For $^1S_0^{[8]}$ and $^3P_J^{[8]}$, the charm-quark fragmentation (Fig.~\ref{fig:qFrag}) is the only production channel. The QCD and QED Feynman diagrams have exactly the same topology and the corresponding amplitudes differ from each other only by the couplings and the color factors. The QCD/QED interference is negative becasue of its negative color factor, and the ratio could be estimated as ${\hat \Gamma}_\mathbb{N}^{{\rm QCD}\times{\rm QED}}/{\hat \Gamma}_\mathbb{N}^{\rm QCD}=-12 q_c^2 \alpha/\alpha_s=-0.171$. 
    \item The $^1P_1^{[8]}$ case is quite similar to the above one, where the charm-quark fragmentation is the most dominant production channel. The only difference is that there exist Fig.~\ref{fig:gFrag} (c, d) diagrams that result in a relatively smaller QCD/QED contribution.
\end{itemize}
Owing to the combination of the larger color factor and the on-shell $Z$ enhancement, the EW corrections from the $HZZ$ diagrams (Fig.~\ref{fig:HZZ}) is also sizable. The relative size of the EW correction is process dependent. The correction for $^1S_0^{[1]}$ is larger than that of $^3S_1^{[1]}$ because the $Zf\bar f$ axial coupling is larger than its vector counterpart. For the color-octet states, the EW corrections are also significant, $\sim 30\%$ of the QCD contributions for $^1S_0^{[8]}$ and $^3P_{J=0,~1}^{[8]}$, and $\sim 140\%$ of the QCD contributions for $^1P_1^{[8]}$ and $^3P_{2}^{[8]}$.  

% This part is for the decay width and BRs.
\begin{table}[tb]
	\caption{The decomposed numerical values of $\Gamma (H \to c  {\bar c} + J/\psi(\eta_c))$ and the corresponding branching fractions. The color-singlet and color-octet contributions are denoted by CS and CO, respectively.}
	\label{tab:SMBodwin}
	\center
	\scalebox{0.9}{
		\begin{tabular}{lccccc}
			\hline
			\multicolumn{1}{c}{}                   & QCD [CS]              & QCD+QED [CS]          & Full [CS]             & Full [CO]             & Full [CS+CO]          \\ \hline
			$\Gamma(H\to c{\bar c}+J/\psi)$ (GeV) & $4.8\times 10^{-8}$ & $5.8\times 10^{-8}$ & $6.1\times 10^{-8}$ & $2.2\times 10^{-8}$ & $8.3\times 10^{-8}$ \\
			${\rm BR}(H\to c{\bar c}+J/\psi)$     & $1.2\times 10^{-5}$ & $1.4\times 10^{-5}$ & $1.5\times 10^{-5}$ & $5.3\times 10^{-6}$ & $2.0\times 10^{-5}$ \\
			\hline
			$\Gamma(H\to c{\bar c}+\eta_c)$ (GeV) & $4.9\times 10^{-8}$ & $5.1\times 10^{-8}$ & $6.3\times 10^{-8}$ & $1.8\times 10^{-7}$ & $2.4\times 10^{-7}$ \\
			${\rm BR}(H\to c{\bar c}+\eta_c)$     & $1.2\times 10^{-5}$ & $1.2\times 10^{-5}$ & $1.5\times 10^{-5}$ & $4.5\times 10^{-5}$ & $6.0\times 10^{-5}$ \\ \hline
		\end{tabular}
	}
\end{table}

For numerical calculations, we employ the $J/\psi$ color-octet long-distance matrix elements (LDMEs) from Ref.~\cite{Bodwin:2014gia}, which is independent of the choice of the color-singlet LDMEs.
Given the SDCs and the LDMEs, it is then straightforward to obtain the decay width $\Gamma (H \to c {\bar c} + J/\psi(\eta_c))$ and the corresponding branching fractions. We decompose the total decay width into color-singlet QCD only, color-singlet QCD+QED, full color-singlet, full color-octet, and full color-singlet plus color-octet and present the numerical results in Table~\ref{tab:SMBodwin}.
The results for the charm-quark fragmentation into color-singlet states are rather robust. In addition,
the QED diagrams introduce a $22\%$ ($2\%$) correction to $J/\psi$ ($\eta_c$) production and the EW correction is $5\%$ ($24\%$) for $J/\psi$ ($\eta_c$).
It is interesting to compare the two mechanisms of the color-singlet and color-octet production.
The production rate of $J/\psi$ ($\eta_c$) through color-octet Fock states is around $36\%$ ($295\%$) of the color-singlet one, which is due mainly to the large $^3S_1^{[8]}$ SDC via the single-gluon fragmentation diagrams. We see that 
the color-octet contribution to $J/\psi$ production is about $1/3$ of the total;
while it is about a factor of 3 larger than the color-singlet contribution for $\eta_c$ production,
because of the large value of $\langle {\cal O}^{\eta_c}[^3S_1^{[8]}] \rangle=\langle {\cal O}^{J/\psi}[^1S_0^{[8]}] \rangle$.
We find it instructive to examine the contributions in some details from different color-octet states as shown in Table~\ref{tab:COBodwin}, where the dominance of $^3S_1^{[8]}$ is shown ($\sim 95\%$ ($100\%$) the total color-octet rate of $J/\psi$ ($\eta_c$) production).
We quote our final results as
\begin{eqnarray}
	&&{\rm BR}(H\to c{\bar c}+J/\psi)= (2.0 \pm 0.5) \times 10^{-5}, 
	\label{eq:BRJ}\\
	&&{\rm BR}(H\to c{\bar c}+\eta_c)= (6.0 \pm 1.0) \times 10^{-5}, 
	\label{eq:BReta}
\end{eqnarray}
where the quoted errors are calculated by using the conservative estimate from the $^3S_1^{[8]}$ LDME fitting as in Ref.~\cite{Bodwin:2014gia}. More work in fitting the LDMEs needs to be done to reduce the uncertainty and improve the precision.  
In comparison with the well-studied decay mode ${\rm BR}(H\to J/\psi + \gamma)=2.8\times 10^{-6}$ \cite{Bodwin:2013gca,Bodwin:2014bpa}, we see an enhancement by an order of magnitude, which is a result of the fragmentation mechanisms.

\begin{figure}[tb]
    \centering
    \includegraphics[width=.48\textwidth]{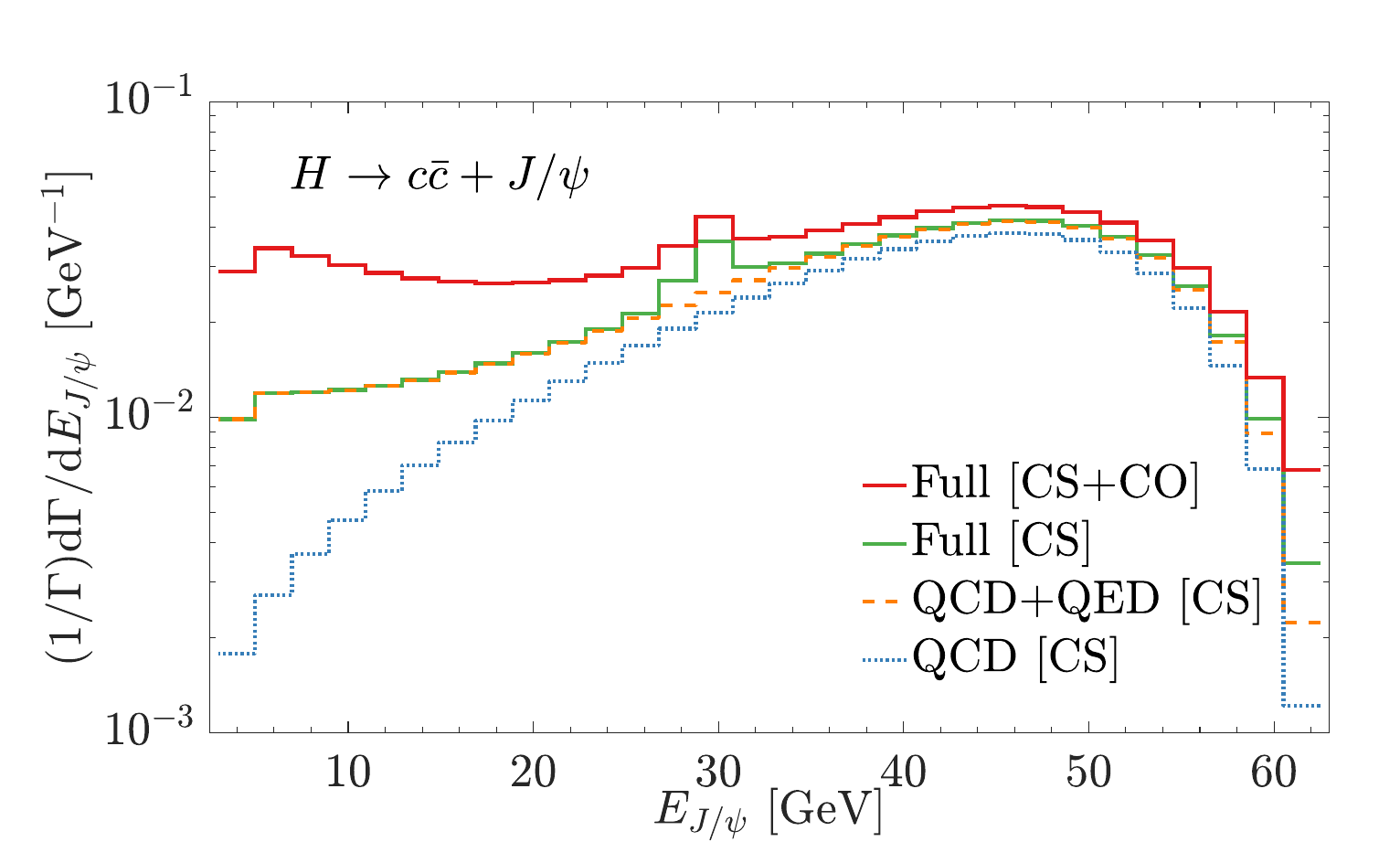}
    \includegraphics[width=.48\textwidth]{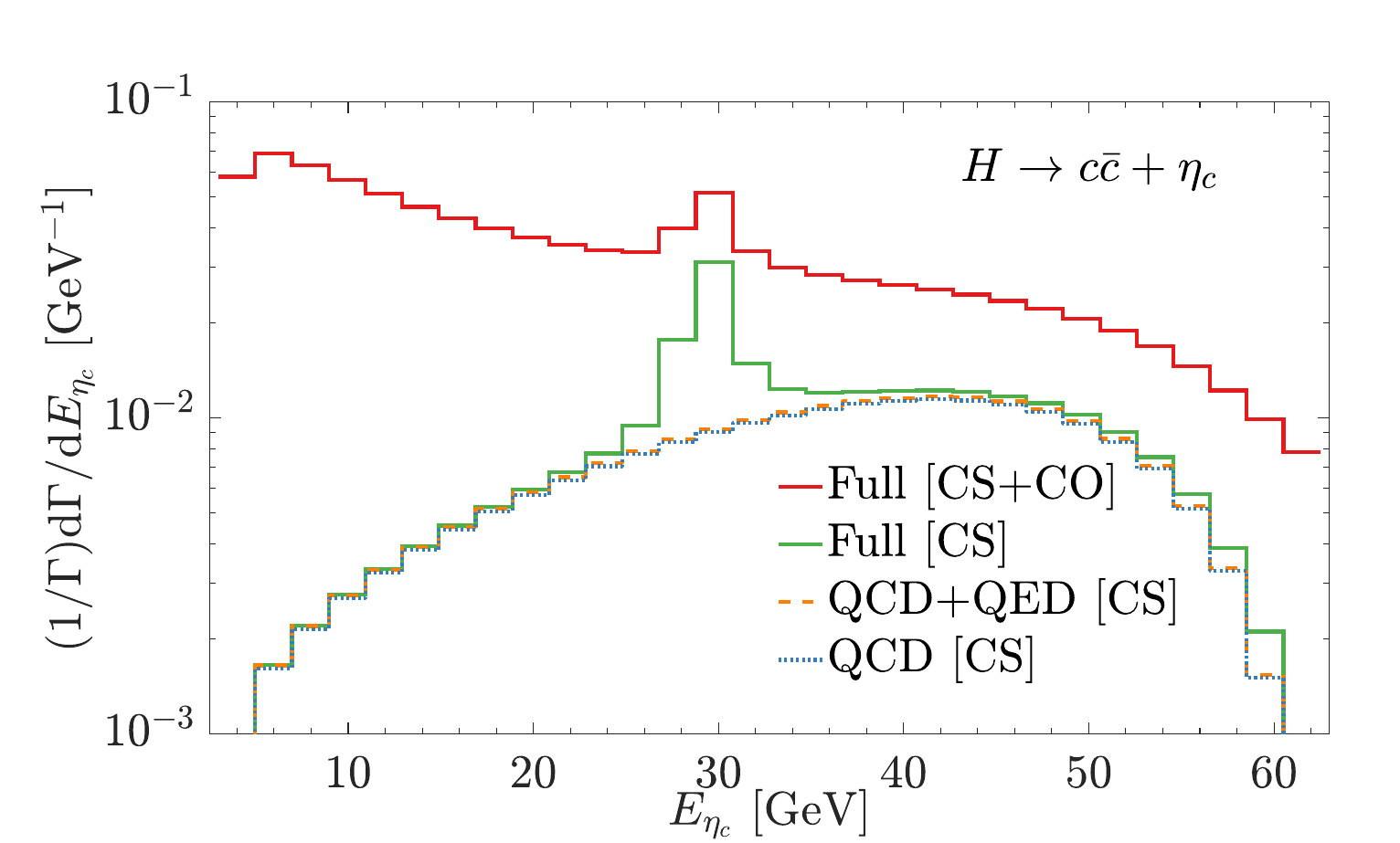}
    \caption{Charmonium energy distributions for (a) $J/\psi$ and (b) $\eta_c$. The blue dotted and orange dashed curves are for color-singlet (CS) QCD only and QCD+QED contributions, the red (green) solid curve is for the sum of full leading order (full CS) result. All curves are normalized using the full leading order decay width in Table \ref{tab:SMBodwin}.} 
    \label{fig:Edis_decom}
\end{figure}

The $J/\psi$ and $\eta_c$ energy distributions $\dd\Gamma/\dd E_{J/\psi(\eta_c)}$ are presented in Fig.~\ref{fig:Edis_decom}. As shown in the plots, the single-photon-fragmentation and single-gluon-fragmentation diagrams have dramatic enhancement on $^3S_1^{[1]}$ and $^3S_1^{[8]}$ production in the low meson energy range, and the charm-quark fragmentation dominates the relative high energy region. As for the EW contribution, it is quite interesting to recognize the enhancements by the approximate two-body kinematics evidenced by the two contributing diagrams as shown in Fig.~\ref{fig:HZZ}: the first diagram yields an on-shell $Z$ process at $E_{J/\psi(\eta_c)} = {1 \over 2} m_H (1-m_Z^2/m_H^2 + 4m_c^2/m_H^2)\approx 30$ GeV; and the second diagram results in a back-to-back kinematics at
$E_{J/\psi(\eta_c)} \approx E_{c\bar c} \approx m_H/2$. These features will serve as an effective discriminator against the contamination from the non-Yukawa contributions.

\begin{figure}[tb]
    \centering
    \includegraphics[width=.48\textwidth]{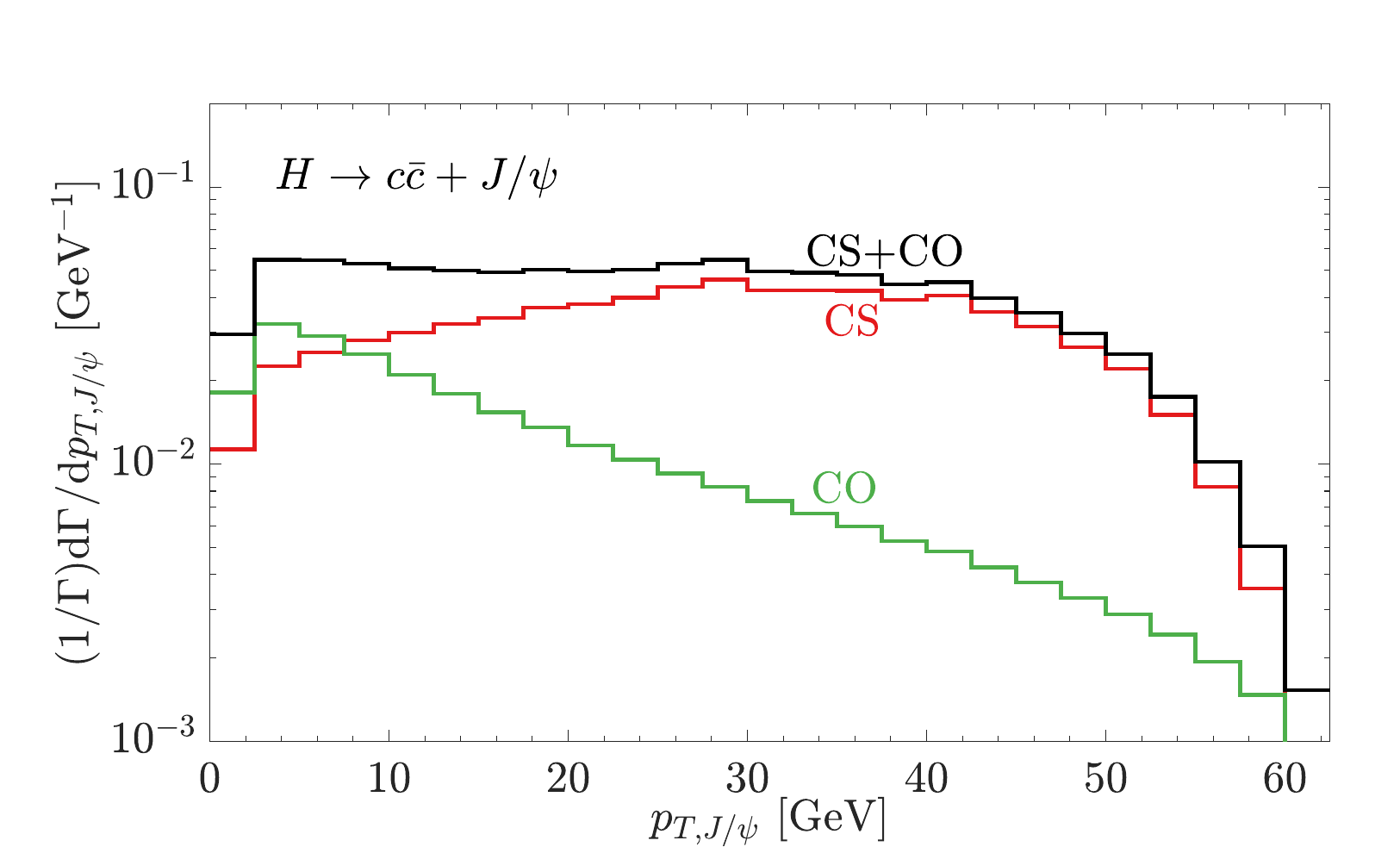}
    \includegraphics[width=.48\textwidth]{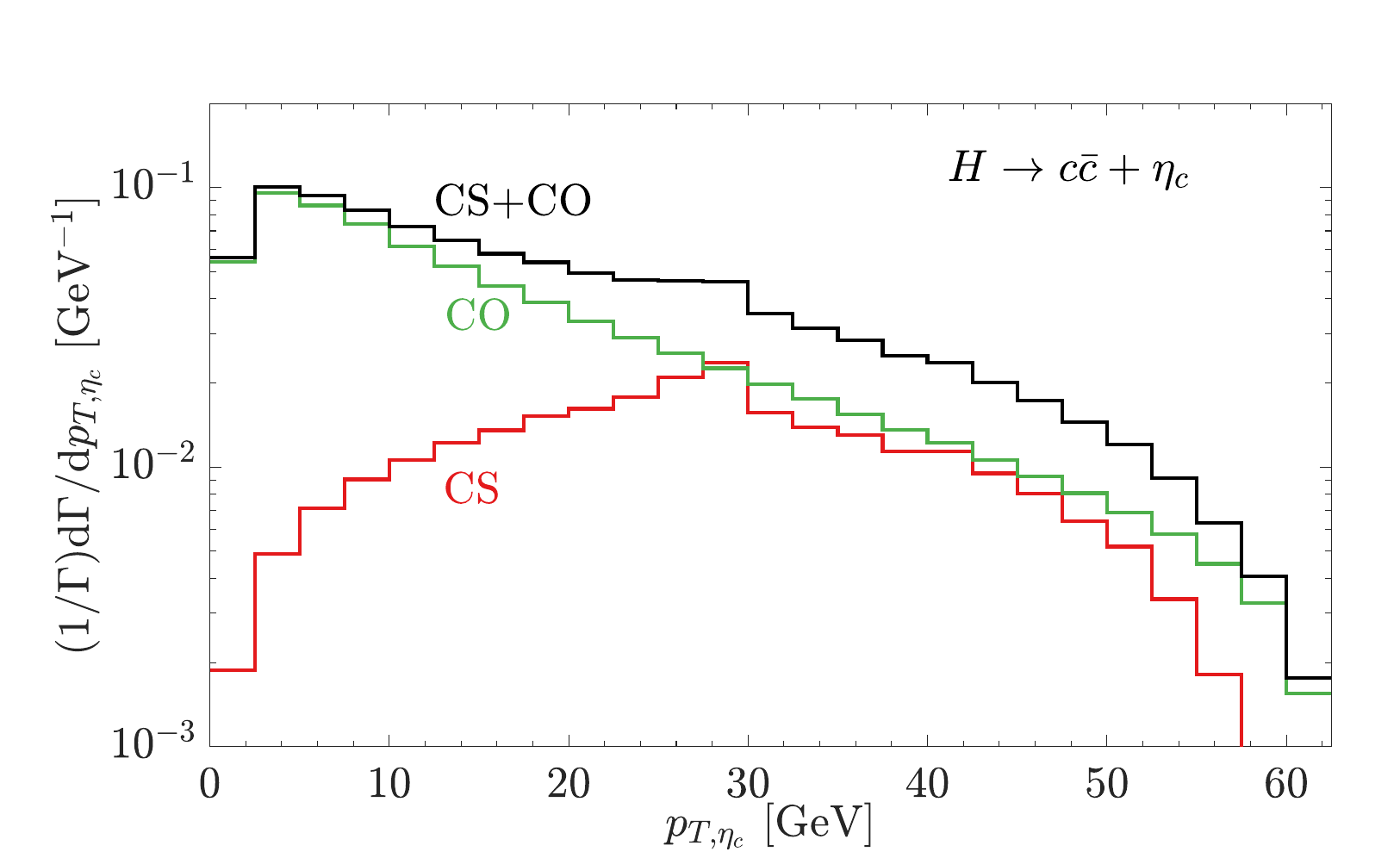}
    \includegraphics[width=.48\textwidth]{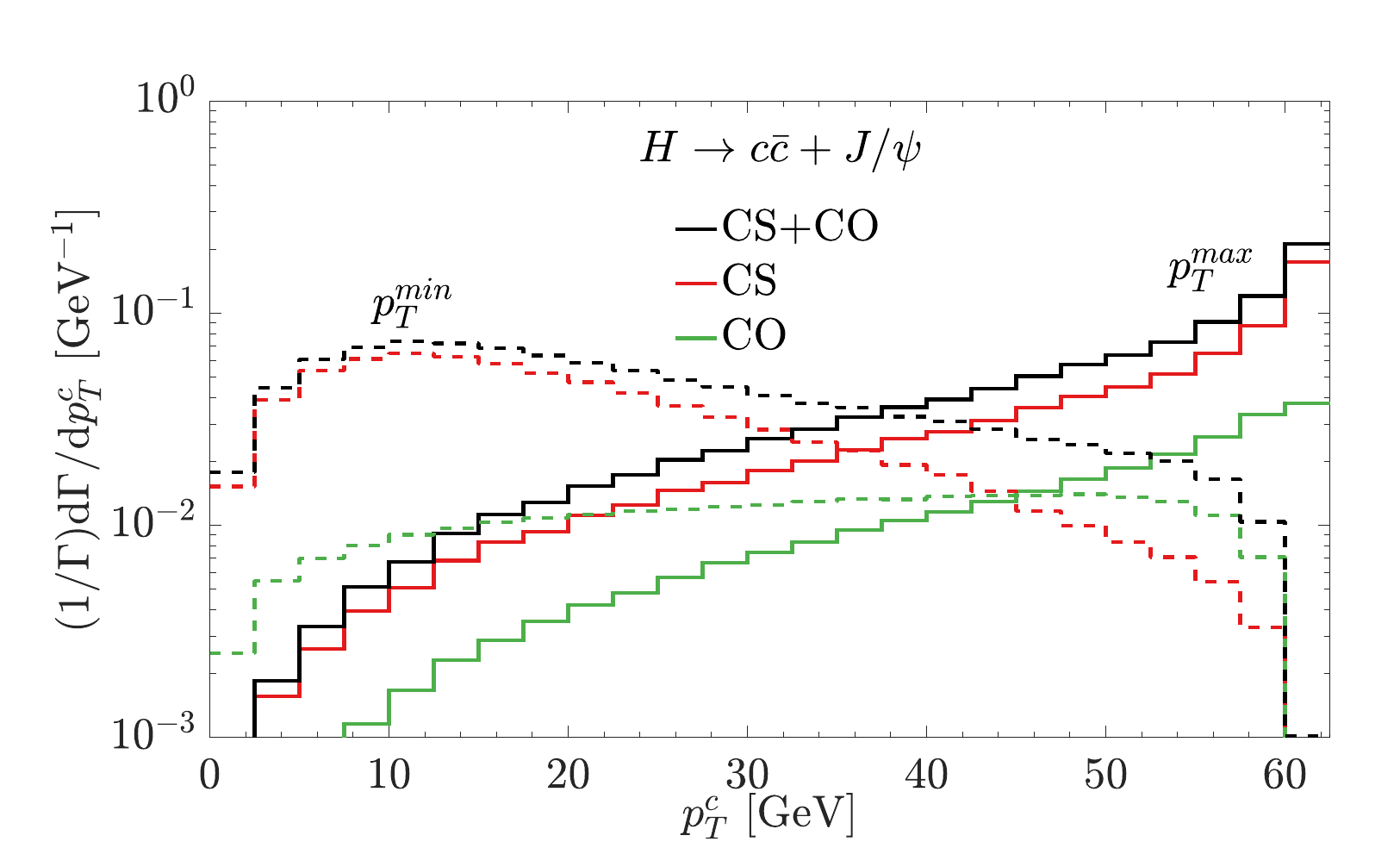}
    \includegraphics[width=.48\textwidth]{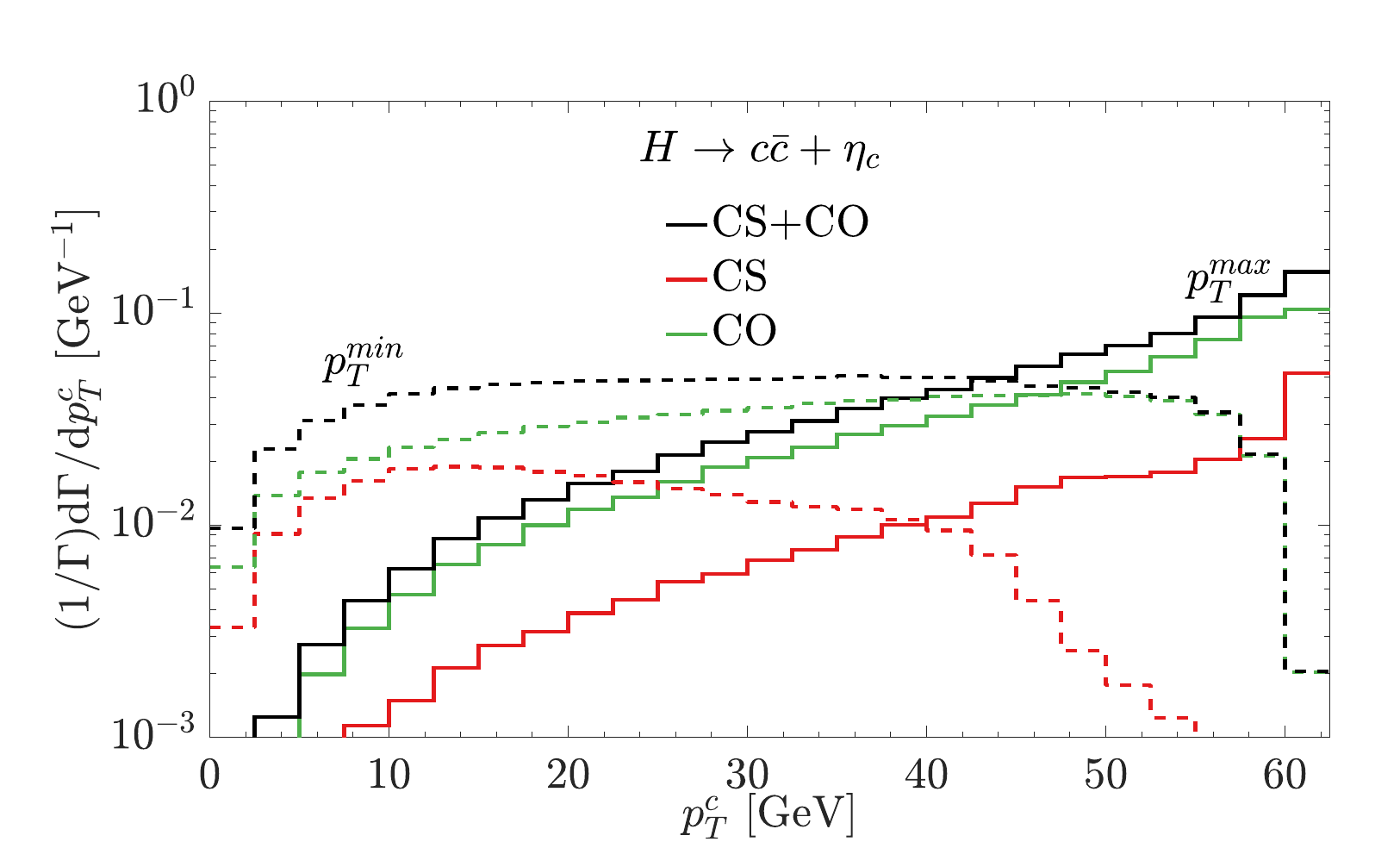}
    \caption{Transverse momentum distributions in the Higgs rest frame $H\to c\bar c+ J/\psi\ (\eta_c)$: 
  (a) and (b) for $J/\psi$ and $\eta_c$ distributions, respectively; (c) and (d) for the charm quark distributions associated with $J/\psi$ and $\eta_c$, respectively, where the solid curves are for the $p_T^{max}$ and dashed curves are for the $p_T^{min}$ distribution. The red, green and black curves are for color-singlet (CS), color-octet (CO), and the full leading order result. All curves are normalized using the full leading order decay width  in Table \ref{tab:SMBodwin}.}
    \label{fig:ptdis}
\end{figure}

From the observational point of view, it is important to predict the transverse momentum spectrum for the decay products. We show the transverse momentum distributions in the Higgs rest frame for $H\to c\bar c+ J/\psi\ (\eta_c)$ in Fig.~\ref{fig:ptdis}: (a) and (b) for $J/\psi$ and $\eta_c$ distributions, respectively; (c) and (d) for the charm quark distributions associated with $J/\psi$ and $\eta_c$, respectively, where the solid curves are for the $p_T^{max}$ and dashed curves are for the $p_T^{min}$ distribution. 
We see that the contribution from the color-octet tends to be softer in $p^{}_{T,J/\psi}$ due to the single-gluon-splitting mechanism, as seen in (a,b); while
the $p_T^{min}$ distribution of the charm quark from the color-singlet tends to be softer, as seen in (c,d), consistent with the fact that the color-signlet mesons are harder.

\begin{table}[tb]
	\caption{The color-octet contributions to $\Gamma (H \to c  {\bar c} + J/\psi(\eta_c))$ and the branching fractions. } \label{tab:COBodwin}
	\center
	\scalebox{0.9}{
		\begin{tabular}{lccccc}
			\hline
			\multicolumn{1}{c}{}                   & $^3S_1^{[8]}$         & $^1S_0^{[8]}$          & $^1P_1^{[8]}$          & $^3P_J^{[8]}$          & Total                 \\ \hline
			$\Gamma(H\to c{\bar c}+J/\psi)$ (GeV) & $2.0\times 10^{-8}$ & $9.8\times 10^{-10}$ & -                      & $2.2\times 10^{-10}$ & $2.2\times 10^{-8}$ \\
			${\rm BR}(H\to c{\bar c}+J/\psi)$     & $5.0\times 10^{-6}$ & $2.4\times 10^{-7}$  & -                      & $5.3\times 10^{-8}$  & $5.3\times 10^{-6}$ \\
			$\Gamma(H\to c{\bar c}+\eta_c)$ (GeV) & $1.8\times 10^{-7}$ & $3.6\times 10^{-11}$ & $1.0\times 10^{-10}$ & -                      & $1.8\times 10^{-7}$ \\
			${\rm BR}(H\to c{\bar c}+\eta_c)$     & $4.5\times 10^{-5}$ & $8.9\times 10^{-9}$  & $2.5\times 10^{-8}$  & -                      & $4.5\times 10^{-5}$ \\ \hline
		\end{tabular}
	}
\end{table}

%%%%%%%%%%%%%%%%%%%%%%%%%%%%%
\subsection{Probing the charm quark Yukawa}

Given the clean decay channels  $J/\psi \to \mu^+\mu^-$ and $e^+e^-$, we will focus on our discussion to the $J/\psi$ mode. With the predicted decay branching fraction of $2\times 10^{-5}$ for $H\to c\bar c + J/\psi$ and 
the Higgs production cross section at the LHC as $ \sigma_H^{}\approx 50$ pb, 
we will expect a signal rate of 1000 event per ab$^{-1}$ integrated luminosity. It is thus promising to search for this channel at the HL-LHC \cite{Apollinari:2017cqg}. We would like to reiterate that the leading contribution to this process directly involves the charm-quark Yukawa coupling, unlike the process $H\to J/\psi+ \gamma$ where the leading contribution is from $\gamma^*\to J/\psi$. 

For simplicity, 
we adopt the $\kappa$ framework~\cite{LHCHiggsCrossSectionWorkingGroup:2013rie} and allow the charm quark Yukawa coupling $y_c$ to deviate from the SM value $y_c^{\rm SM}$ by a factor of $\kappa_c$
\begin{eqnarray}
    y_c=\kappa_c y_c^{\rm SM}.
\end{eqnarray}
Neglecting the sub-leading contributions from the EW and top-loop diagrams, the branching fractions and thus the production rates for the processes under consideration scale with the charm-Yukawa coupling as
\begin{eqnarray}
    {\rm BR} \approx \kappa_c^2\ {\rm BR}^{\rm SM}.
    \label{eq:BR}
\end{eqnarray}
Assuming a detection efficiency $\epsilon$ for the final state $c\bar c+\ell^+\ell^-\ (\ell=\mu,e)$ and an integrated luminosity $L$, we write the anticipated number of events as 
\begin{equation}
N = L \sigma_H^{}\ \epsilon\ {\rm BR}(c\bar c+\ell^+\ell^-) \approx 12\ \kappa_c^2 \times {L\over {\rm ab}^{-1}}\times {\epsilon \over 10\%}, 
\end{equation}
where the 12\% branching fraction for $J/\psi \to \mu^+\mu^-, e^+e^-$ has been included.\footnote{The 
$10\%$ efficiency for $\epsilon$ is a rough estimate with a double charm tagging of $(40\%)^2$ \cite{Aaboud:2018fhh}, and a kinematic acceptance of $50\%$ based on the distributions in Fig.~\ref{fig:ptdis}.} 
Considering the statistical error only $\delta N \sim \sqrt{N}$, one would get an accuracy for the coupling determination
\begin{equation}
\Delta\kappa_c \approx 15\% \times ({L\over {\rm ab}^{-1}}\times {\epsilon \over 10\%})^{-1/2}.  \label{eq:Deltakappac}
\end{equation}

\begin{figure}[tb]
    \centering
    \includegraphics[width=.48\textwidth]{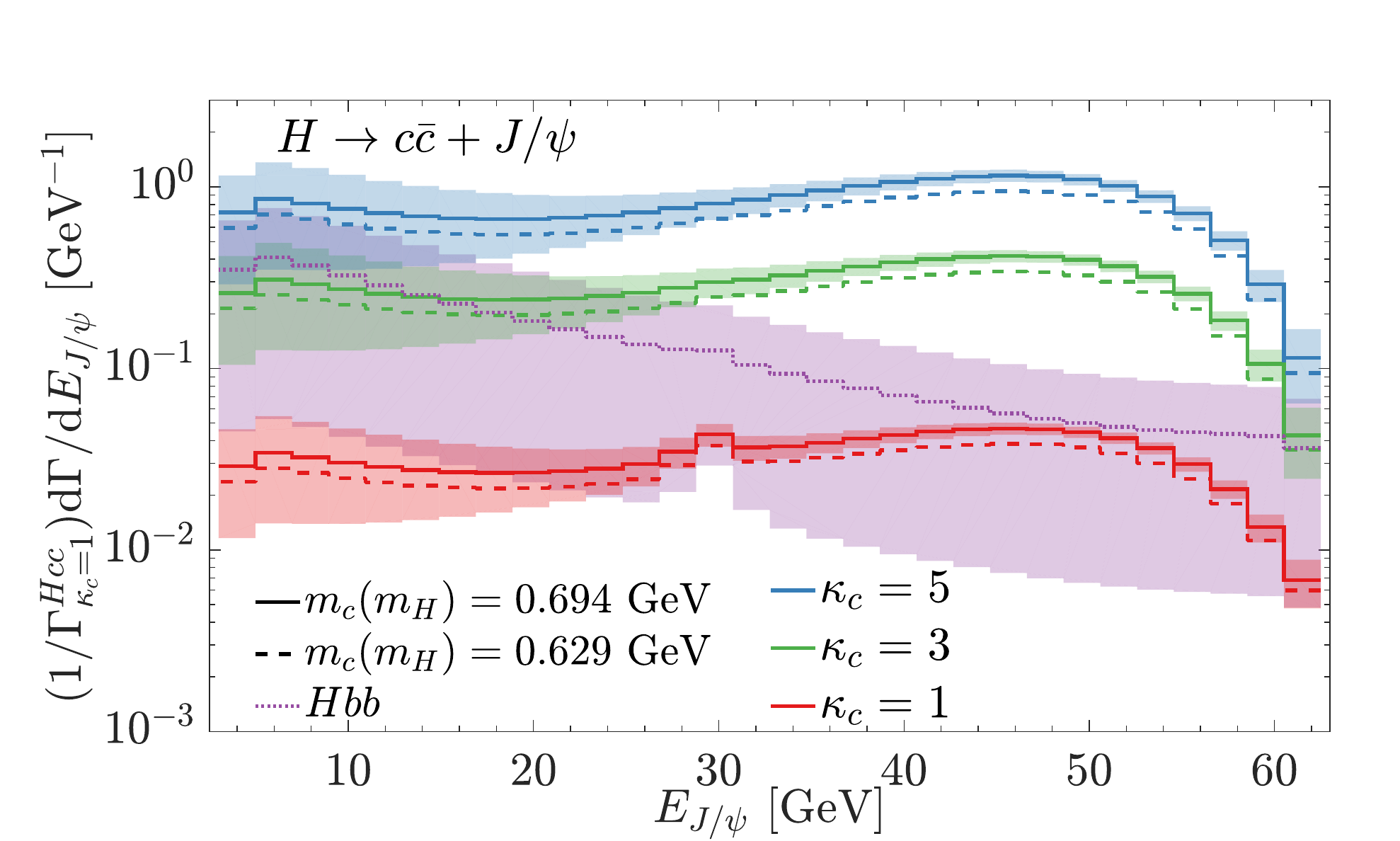}
    \includegraphics[width=.48\textwidth]{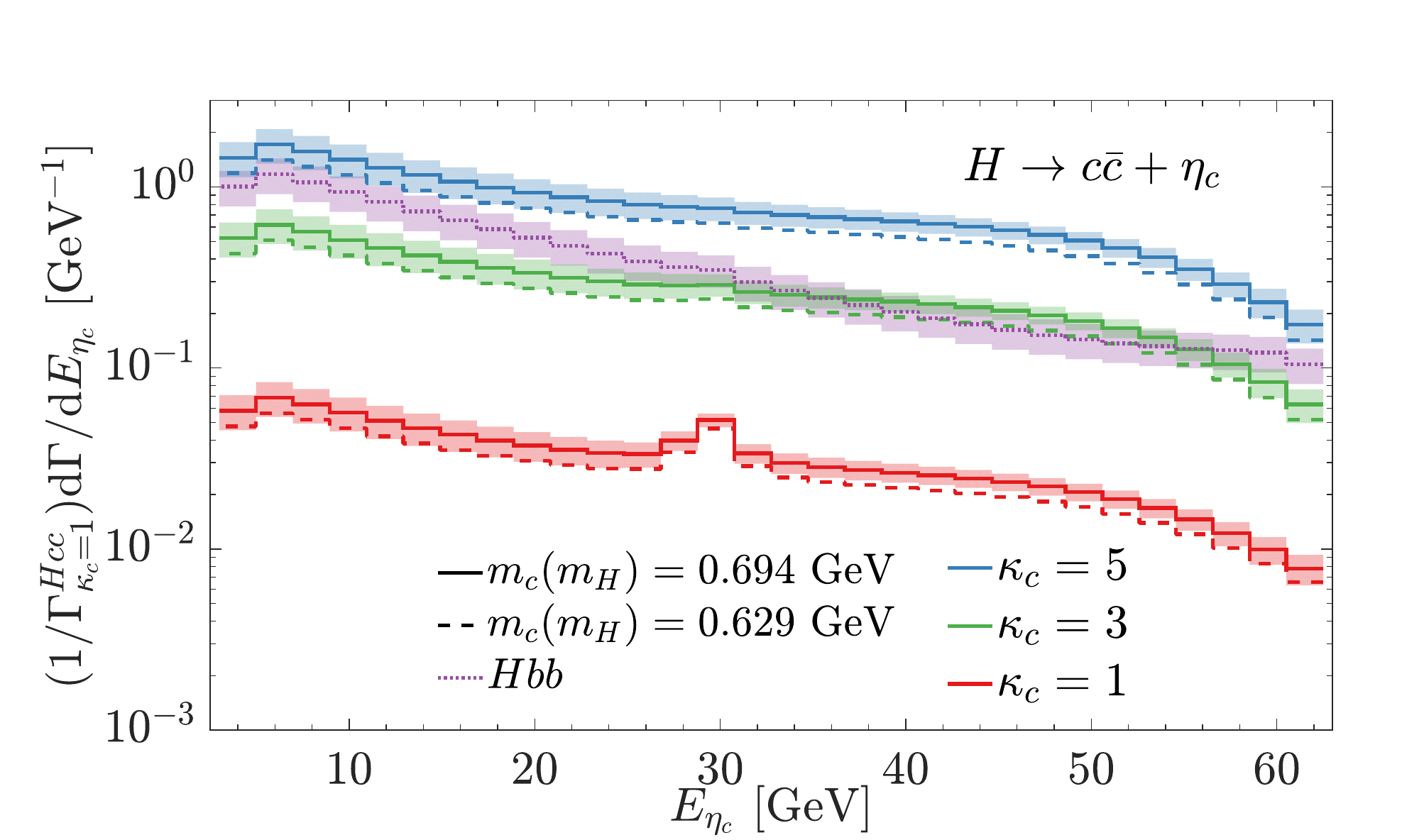}
    \caption{Charmonium energy distributions for (a) $J/\psi$ and (b) $\eta_c$ for $\kappa_c=1~({\rm SM}),\,3,\,5$. The solid curves are for the one-loop running mass $m_c(m_H)=0.694~{\rm GeV}$; the dashed curves are for the four-loop running mass $m_c(m_H)=0.629~{\rm GeV}$. 
    The dotted purple curve is for the background from $Hbb$ decay mode. The colored bands are for the uncertainties from the color-octet long-distance matrix elements (LDMEs). All curves are normalized using the full  SM leading order decay width in Table \ref{tab:SMBodwin}.} 
    \label{fig:Edis_k_Hbb}
\end{figure}

In Fig.~\ref{fig:Edis_k_Hbb}, we show the $J/\psi$ and $\eta_c$ energy distributions for a few illustrative couplings $\kappa_c=1~({\rm SM}),\,3,\,5$, by the red, green and blue curves, respectively. 
We note that, the results confirm the simple, yet important, relation in Eq.~(\ref{eq:BR}). The EW contribution is seen near $E_{J/\psi(\eta_c)} \approx 30$ GeV, that does not follow this relation, and it becomes invisible for larger values of $\kappa_c$.  
To have a more complete comparison, we also employ the four-loop charm quark running mass $m_c(m_H)=0.629~{\rm GeV}$ via the package RunDec \cite{Chetyrkin:2000yt,Herren:2017osy} in addition to the one-loop $m_c(m_H)=0.694~{\rm GeV}$, as shown by the dashed curves correspondingly. 
The color-octet long-distance matrix element (LDME) uncertainties are indicated by the colored bands around the solid curves. We see that the uncertainty is more significant at the low energy region, due to the enhancement of the single-gluon-fragmentation diagrams to the $^3S_1^{[8]}$ contribution which has a large uncertainty. 

%%%%%%%%%%%%%%%%%%%%%%%%%%%%%%%%%%%%
\subsection{Backgrounds}
In the realistic experimental search for the signal $H\to c\bar c + J/\psi$ at the LHC, there are large backgrounds to the signal. 

The quarkonium production mechanism could result in hadronic jets associated with the quarkonium state, which serve as the main background for the signal $H\to c\bar c + J/\psi$.
The formidable background is the associated production of $J/\psi$ and light $g,q$-jets. The cross section of the prompt $J/\psi$ production has been measured to be ${\rm BR}(J/\psi \to \mu^+\mu^-)\times\sigma(pp\to J/\psi)\simeq 860~$pb for $20\leq p_T\leq 150~{\rm GeV}$, with a data sample of $2.3~{\rm fb}^{-1}$ by CMS \cite{CMS:2017dju},\footnote{We obtained the cross section by summing over the data from their differential cross section.}
where the $J/\psi$ state is reconstructed in the dimuon decay channel for dimuon rapidity $|y|<1.2$.
Requiring to tag two additional charm-like jets from the inclusive $J/\psi$ sample would likely reduce this background rate by several orders of magnitude. Detailed simulation would be needed for charm tagging and kinematic optimization in order to draw a quantitative conclusion for the signal observability.

The leading irreducible background comes from the QCD production of $J/\psi$ plus heavy flavor jets, {\it i.e.} $gg, q\bar q \to c\bar c + J/\psi$. As presented in Ref.~\cite{Artoisenet:2007xi}, the cross section falls sharply versus the transverse momentum,
dropping by $4$ orders of magnitude at $p_T\simeq 20~$GeV.
Experimental measurements of such processes at the LHC have not yet been performed, but the high performance of jet flavor tagging at ATLAS/CMS offers potential for the future measurements at the HL-LHC \cite{Chapon:2020heu}. The event yield was estimated to be 75000 with an integrated luminosity of 3 ab$^{-1}$ \cite{Chapon:2020heu}, translating to a cross section of 25 fb. 
Although this background rate is large comparing with the expected signal about 1 fb, their kinematical distributions are quite different from the Higgs decay. We may expect to reduce the background by applying some suitable judicious kinematic cuts.

\begin{figure}[tb]
	\centering
	\includegraphics[width=.24\textwidth]{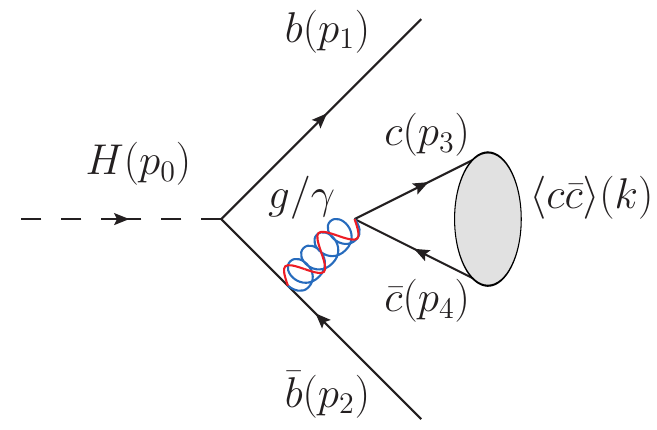}
	\includegraphics[width=.24\textwidth]{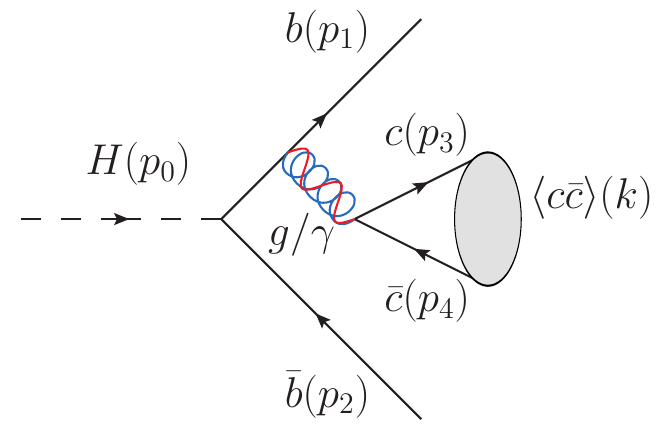}
	\includegraphics[width=.24\textwidth]{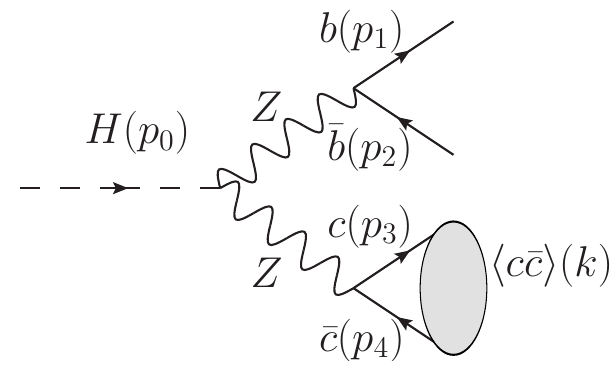}
	\includegraphics[width=.24\textwidth]{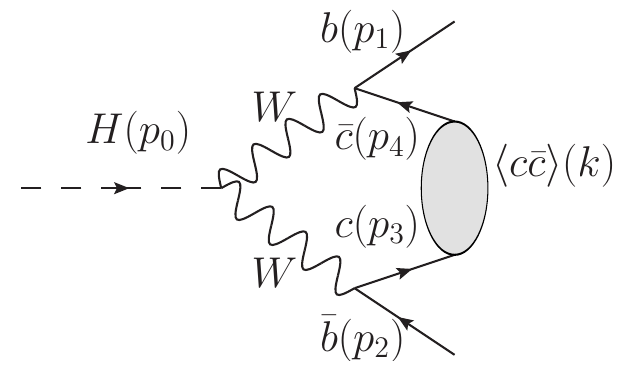}
	 \caption{Feynman diagrams for $ H\to b\bar{b}+ J/\psi (\eta_c)$ production. The gluon diagrams in (a, b) only contribute to $ ^3 S_1 ^{[8] }$, while the photon ones only contribute to $ ^3 S_1 ^{[1]} $. (c) is nonzero only for the color-singlet states.}\label{fig:Hbb}
\end{figure}

In addition, due to the larger $Hb \bar b$ coupling, the decay $H\to b{\bar b}+J/\psi(\eta_c)$, as shown in Fig.~\ref{fig:Hbb}, may yield significant contamination to the test of the charm-Yukawa coupling. Following our calculational formalism, it is straightforward to obtain the corresponding branching fractions as 
\begin{eqnarray}
    &&{\rm BR}(H\to b{\bar b}+J/\psi)= (8.6 \pm 7.5) \times 10^{-5}, \nonumber \\
    &&{\rm BR}(H\to b{\bar b}+\eta_c)= (7.4 \pm 1.6) \times 10^{-4}, 
    \label{eq:bb}
\end{eqnarray}
where the main contribution is through the $^3S_1^{[8]}$ state and the errors are estimated using the uncertainty of $\langle {\cal O}^{J/\psi}[^3S_1^{[8]}] \rangle$. We note the large uncertainty which is attributed to both the large $Hbb$ coupling and the single-gluon-fragmentation enhancement. To appreciate the relative size, 
we present the charmonium energy distributions from $b\bar b+ J/\psi\ (\eta_c)$ productions in Fig.~\ref{fig:Edis_k_Hbb}, as shown by the purple curve. The band around it indicates the uncertainty.\footnote{The $H\to b\bar b+ J/\psi$ was recently calculated to NLO in $\alpha_s$ \cite{Pan:2022nxc}; our estimation is consistent with their LO results under the same parameter settings.} 
Its overall rate is about a factor of 4 larger than that of $c\bar c + J/\psi$. It is quite conceivable that an effective charm-tagging would be implemented to separate those two contributions in the experimental analysis.

%%%%%%%%%%%%%%%%%%%%%%%%%%%%%%%%%%%
\section{Summary}
It is of fundamental importance to study the Higgs boson couplings to light fermions. It is extremely challenging to test the charm-quark Yukawa coupling at hadron colliders due to the large QCD background to the decay $H\to c\bar c$. 
Instead, other decay modes that may be sensitive to the coupling have been suggested. 
In this paper, we considered a new decay channel, the Higgs boson decay to $J/\psi$ and $\eta_c$ states via the charm-quark fragmentation. 
We calculated the branching fractions for the decays in Eq.~(\ref{eq:HtoJ}). 
The decay rates are governed by the charm-quark Yukawa coupling $y_c$, unlike the decay $H\to J/\psi + \gamma$, which is dominated by the $\gamma^*$-$J/\psi$ mixing. 

We performed the calculation in the framework of NRQCD, including the contributions of both the color-singlet and color-octet mechanisms, as well as the electroweak contributions from the $HZZ$ coupling. For the color-singlet production, we adopted the long-distance matrix elements (LDMEs) from the most updated value of wave function at origin $R(0)$ \cite{Eichten:2019hbb}.
For the color-octet production mechanism, the LDMEs would have to be extracted from fitting the experimental data. There is a significant uncertainty from the fitting \cite{Bodwin:2014gia,Chao:2012iv,Feng:2018ukp}, and we adopt the results from Ref.~\cite{Bodwin:2014gia}, which used the high $p_T$ data as input, more relevant to the situation of our current consideration.

It is interesting to note the different relative sizes from contributions of color-singlet versus color-octet.
Numerically, we found that the contribution from the color-singlet state is about three times larger for $J/\psi$ (but three times smaller for $\eta_c$) than that from the color-octet states. 
We found the electroweak contributions to the decay via the $HZZ$ coupling to be small for $^3S_1^{[1]}$ (the color-singlet contribution to $J/\psi$) and $^3S_1^{[8]}$ (the main color-octet contribution to both $J/\psi$ and $\eta_c$), at the order of percentage. 
We finally commented on the sub-leading contributions from $H\to g^*g^*$ 
via the top-quark loop, and from $H\to \gamma^*\gamma^*$ via the top-quark and $W$ loops. 
We conclude that the decay branching fractions are
\begin{eqnarray}
	{\rm BR}(H\to c{\bar c}+J/\psi)\approx 2.0 \times 10^{-5}\quad {\rm and}\quad
	{\rm BR}(H\to c{\bar c}+\eta_c) \approx 6.0 \times 10^{-5}.
\end{eqnarray}

We comment on the perspective on searching for the Higgs to $J/\psi$ transition at the HL-LHC for testing the charm-quark Yukawa coupling. 
If only based on the statistics, with the large Higgs boson production rate anticipated at the HL-LHC of 50 million per ab$^{-1}$, we would expect to reach a sensitivity of about $15\%$ on the coupling $y_c$, which is in the same ballpark as the $\sim 25\%$ theoretical uncertainty in Eq.~(\ref{eq:BRJ}) and the $\sim 16\%$ EW contamination from the $HZZ$ ($3\%$) and the $H\to g^*g^*/\gamma^*\gamma^*$ ($13\%$) channels. There are, however, formidable SM QCD backgrounds for this channel. Assuming $10,000$ background events after the selection cuts at the HL-LHC, one could reach a $2\sigma$ sensitivity for the coupling $\kappa_c\approx 2.4$.  
Detail analyses including the detector and the systematic effects would be called for to reach a quantitative conclusion. 

Our formalism and results are also applicable to the Higgs decays to other fragmentation channels with heavy quarkonia, if the heavy quark mass is properly adjusted, as explicitly shown for 
$H\to b\bar b+J/\psi\ (\eta_c)$ in Eq.~(\ref{eq:bb}).

%%%%%%%%%%%%%%%%%%%%%%%%%%
\acknowledgments
We thank Florian Herren, Yongcheng Wu for helpful discussions, and Satya Mukhopadhyay for collaboration at the initial stage of the project. 
This work was supported in part by the U.S.~Department of Energy under grant 
No.~DE-SC0007914, 
U.S.~National Science Foundation under Grant No.~PHY-2112829, and in part by the PITT PACC. 
The support provided by China Scholarship Council (CSC) during the visit of Xiao-Ze Tan to PITT PACC is also acknowledged.
  
%%%%%%%%%%%%%%%%%%%%%%%%%%%%%
\newpage
\section*{Appendix}
\appendix
\section{Polarization sum} \label{Appendix:polsum}
In order to perform the proper polarization sums, we define
\begin{eqnarray}
    \Pi_{\alpha\beta} \equiv -g_{\alpha\beta}+\frac{P_\alpha P_\beta}{m^2},
\end{eqnarray}
where $m=2 m_c$ is the mass of the $Q{\bar Q}$ bound state.
\begin{itemize}
    \item For $^3S_1$ and $^1P_1$ states, the polarization sum is 
    \begin{eqnarray}
        \sum_{h} \epsilon_\alpha \epsilon^*_{\alpha'} =\Pi_{\alpha \alpha'},
    \end{eqnarray}
    \item For $^3P_J$ states, there are three multiplets, {\it i.e.} $J=0,\,1,\,2$. We need to define the polarization tensor ${\cal E}_{\alpha\beta}^{(J)}$
    \begin{eqnarray}
        &&{\cal E}_{\alpha\beta}^{(0)} {\cal E}_{\alpha'\beta'}^{(0)*}=\frac{1}{3} \Pi_{\alpha \beta} \Pi_{\alpha' \beta'}\nonumber, \\
        &&{\cal E}_{\alpha\beta}^{(1)} {\cal E}_{\alpha'\beta'}^{(1)*}=\frac{1}{2} \left(\Pi_{\alpha \alpha'} \Pi_{\beta \beta'}-\Pi_{\alpha \beta'} \Pi_{\alpha' \beta}\right)\nonumber, \\
        &&{\cal E}_{\alpha\beta}^{(2)} {\cal E}_{\alpha'\beta'}^{(2)*}=\frac{1}{2} \left(\Pi_{\alpha \alpha'} \Pi_{\beta \beta'}+\Pi_{\alpha \beta'} \Pi_{\alpha' \beta}\right)-\frac{1}{3} \Pi_{\alpha \beta} \Pi_{\alpha' \beta'}.
    \end{eqnarray}
\end{itemize}

%\nocite{*}
% \bibliographystyle{utphys}
\bibliographystyle{JHEP}
\bibliography{ref.bib}

\end{document}